\documentclass[a4paper,fleqn,usenatbib,useAMS]{mnras}

\usepackage{graphicx}	
\usepackage{amsmath}	
\usepackage{amssymb}	
\usepackage{multicol}        
\usepackage{bm}		
\usepackage{pdflscape}	
\usepackage{journals}

\newcommand{\kmps}{km~s\ensuremath{^{-1}}\,}
\newcommand{\Msun}{M\ensuremath{_\odot}\,}

\newcommand{\Oo}{\displaystyle}

\newcommand{\sk}[1]{\textcolor[rgb]{0.20,0.80,0.00}}

\usepackage[T1]{fontenc}
\usepackage{ae,aecompl}
\usepackage{epstopdf}	
\usepackage{newtxtext,newtxmath}

\title[The study of two barred galaxies]{The study of two barred galaxies with curious kinematical features\thanks{ Based on observations obtained with the 6-m telescope of the
Special Astrophysical Observatory of the Russian Academy of
Sciences (SAO RAS).}}
\author[A. Saburova et al.]{A. S. Saburova$^{1}$\thanks{E-mail:
saburovaann@gmail.com}, I. Yu. Katkov$^{1}$,  S. A. Khoperskov$^{2,3}$, A. V. Zasov$^{1,4}$\thanks{E-mail:
zasov@sai.msu.ru}, R. I. Uklein$^{5}$ 
\\
$^1$ Sternberg Astronomical Institute, Moscow M.V. Lomonosov State University, Universitetskij pr., 13,  Moscow, 119991, Russia\\
$^2$ GEPI, Observatoire de Paris, PSL Research University, CNRS, Place Jules Janssen, 92190 Meudon, France \\
$^3$ Institute of Astronomy, Russian Academy of Sciences, 48 Pyatnitskaya st., 119017 Moscow, Russia\\
$^4$ Faculty of Physics, Moscow M.V. Lomonosov State University, Leninskie gory 1,  Moscow, 119991, Russia \\
$^5$ Special Astrophysical Observatory, Russian Academy of Sciences, Nizhniy Arkhyz, Karachai-Cherkessian Republic 357147, Russia \\
}

\begin{document}
\label{firstpage}
\pagerange{\pageref{firstpage}--\pageref{lastpage}} \pubyear{2016}
\maketitle

\begin{abstract}
We performed long-slit spectral observations of two SB-type galaxies: NGC~5347, UGC~1344. They were previously
suspected as the galaxies with unusually low mass-to-light ratios (on the ground of mass estimates from the H{\sc i} linewidths), which are in conflict with their observed colours. The observations were conducted at the Russian 6-m telescope. The aim of the study was to clarify the kinematics and structure, as well as the properties of stellar populations of the galaxies. The results of observations disproved the peculiarly low mass-to-light ratios of both galaxies. The most probable reasons of underestimation of their masses are discussed. We tried to reproduce the main observed features of kinematical profiles of the galaxies in the $N$-body simulations of barred galaxies. 
We found that both galaxies possess central components of different structure. Indeed, the age and velocity dispersion of stellar population in NGC~5347 are low in its innermost part in comparison to that of the bulge or the bar, which agrees with the presence of nuclear kinematically decoupled disc. It probably was formed due to  the bar which supplied the inner region with  gas. The kinematical profiles of the second galaxy UGC~1344 give evidences in favour of the central peanut-shaped bulge. In spite of the different luminosities of the two galaxies, they possess nearly equal (close to solar) central stellar abundance and the flattening of the stellar metallicity gradient in the bar regions. However, in the less luminous NGC~5347 the mean stellar age is younger than that in UGC~1344.
\end{abstract}

\section{Introduction}\label{intro}

Stellar initial mass function (IMF) has a fundamental meaning to understand the evolution of galaxies and star formation. However, the question of its universality  still remains a matter of hot discussion. There are evidences either in favor or against it (see e.g. \citealt{Kroupa2002}; \citealt{Gilmore}; \citealt{Bastian}; \citealt{Davisetal2016}). In \cite{Saburova2009} we tried to find the candidates of discy galaxies where one can suspect a significant difference of stellar IMF from the ``standard'' Salpeter  IMF \citep{Salpeter1955} or Kroupa \citep{Kroupa2001} one. We were looking for such galaxies among the objects with peculiarly low or high dynamical mass-to-light ratios $M/L_B= v^2r_{25}/\left(GL_B\right)$ which conflict with observed colour indices, where $v$ is the velocity of rotation,  $r_{25}$ is the optical radius and $L_B$ is the B-band total luminosity. We studied in details several galaxies suspected for unusual M/L ratios of their stellar populations in subsequent papers: \cite{Saburova2015}, \cite{saburovaetal2013}, \cite{Saburovaetal2011}. For the candidates we considered  in these papers the new photometric and kinematic data have not confirmed the expected peculiarity, giving evidences in favour of the universality of the IMF in discy galaxies. However, we continue our study of the galaxies suspected with peculiar mass-to-light ratios. 

The aim of the current paper is the study of two galaxies from our list of galaxies with presumably unusually low $M/L_B$  ratio on the basis of Hyperleda\footnote{http://leda.univ-lyon1.fr/}  catalogue data (\citealt{Makarov2014}):  ~NGC~5347 and UGC~1344. Their main properties are presented in Table \ref{properties}. Both are early-type disc galaxies, possessing similar red colours of stellar population, a ring and a prominent bar. The luminosity of UGC~1344 is significantly higher than that of NGC~5347. The inner ring of NGC~5347  is distinguished by its FUV continuum and is conspicuous in the continuum-subtracted $\rm H \alpha$ images  which indicates the active star formation there (\citealt{Comeron2013}). The ring of UGC~1344 is not perceptible in FUV, hence it should be also passive in $\rm H \alpha$~(\citealt{Comeron2013}), being a fossil ring not hosting star formation.
  NGC~5347 is a Seyfert 2 galaxy with the mass of central black hole of $\log(M_{\rm bh}/\Msun)=8.73$ (\citealt{Izumi2016}). The H{\sc i} observations of NGC~5347 demonstrate that it has a peculiar morphology: the gaseous disc position angle is almost perpendicular to that of the optical disc. There is a hint that the H{\sc i} is elongated along the major axis of the optical disc in the inner part of the galaxy, in this case the galaxy may have a strong warp,  however new H{\sc i}  observations with better resolution are needed to confirm or disprove it (\citealt{Noordermeer2005}). 
Another interesting detail considering this galaxy is that according to HST observations it possesses a very well defined   nuclear dust spiral structure (\citealt{Martini2003}). This finding is in good agreement with the kinematical properties obtained in the current article (see below). It seems unlikely that the formation of NGC~5347 was affected by recent interaction. According to \cite{Garcia-Barreto2003} NGC~5347 does not have companions within the distance of its $10-20$ diameters, however it belongs to Canes Venatici Camelopardalis cloud (\citealt{Tully1987}). UGC~1344 belongs to a group (\citealt{Garcia1993}) and,  as we will show below, has a small companion 2MASX J01522797+3629533 at the distance of roughly 24 kpc.

As it will be demonstrated in this paper, the peculiarity of the mass-to-light ratios of the considered galaxies did not find a confirmation by the new kinematical data. Nevertheless, these systems represent the interesting cases due to the curious inner kinematical features and deserve the detailed study we have performed. 

The current paper is organized as follows: Section \ref{Obs} is devoted to the details of data reduction and the results of observations; the discussion including the comparison with our $N$-body simulations is given in Section \ref{Discussion}; the main results are summarized in Section \ref{conclusion}. 

\begin{table} 
\caption{Main properties of the observed galaxies: name; adopted distance; morphological type; apparent $B$-band stellar magnitude corrected for the extinction; inclination angle; $B-V$ colour index corrected for the  extinction }\label{properties}
\begin{center}
\begin{tabular}{cccccc}
\hline\hline
Galaxy&D&type&$m_B$&$i$&$(B-V)_0$ \\
   &(Mpc)&&(mag)&(\degr) &(mag)  \\
\hline
NGC~5347&32& SBab&13.16&45&0.7 \\
UGC~1344&58& SBa&12.98&61&0.74 \\
\hline\hline
\end{tabular}
\end{center}
\end{table}

\section{Observations }\label{Obs}
\subsection{Reduction of spectral observations}
\begin{figure*} 
\centering
\includegraphics[height=0.32\textheight]{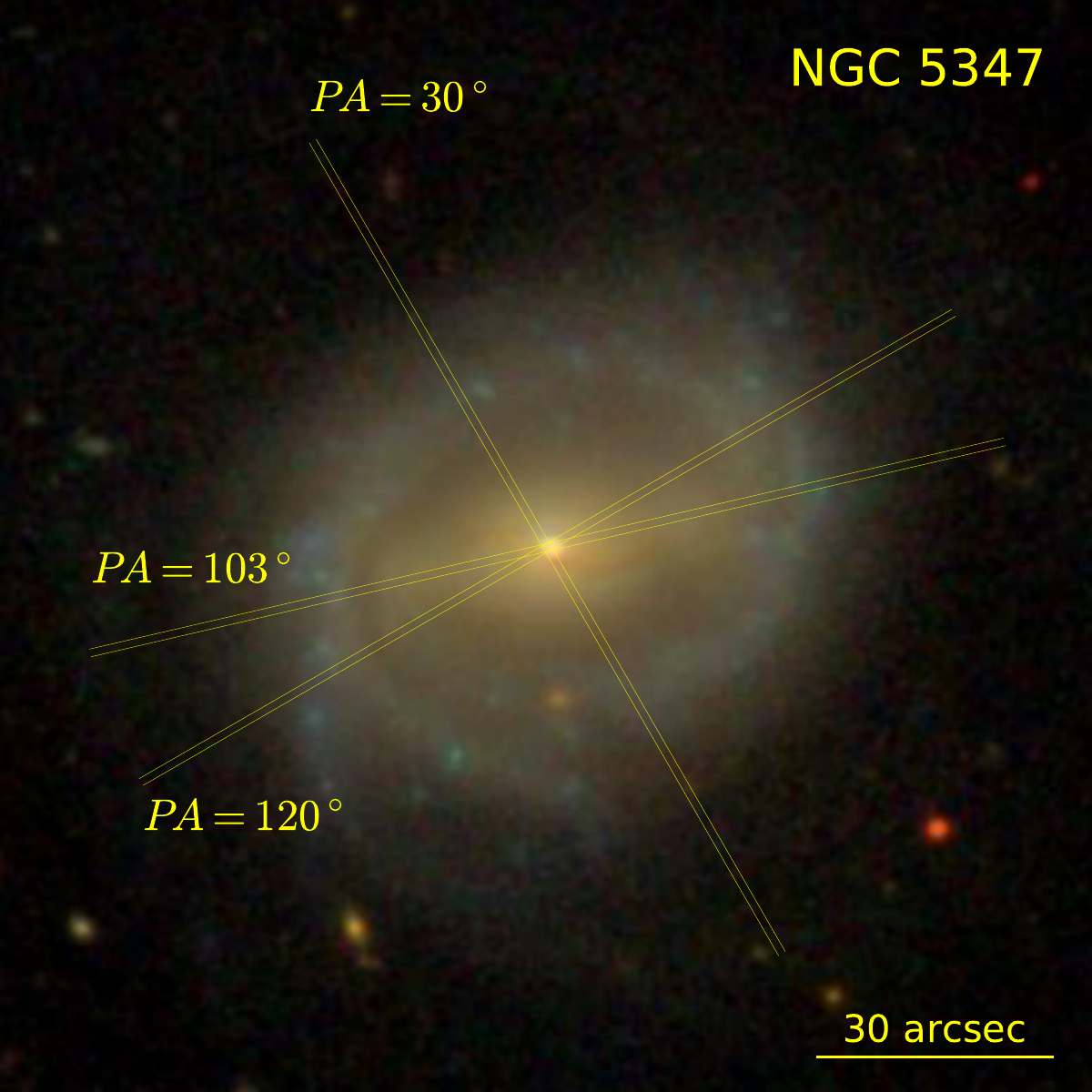}
\hspace{0.5cm}
\includegraphics[height=0.32\textheight]{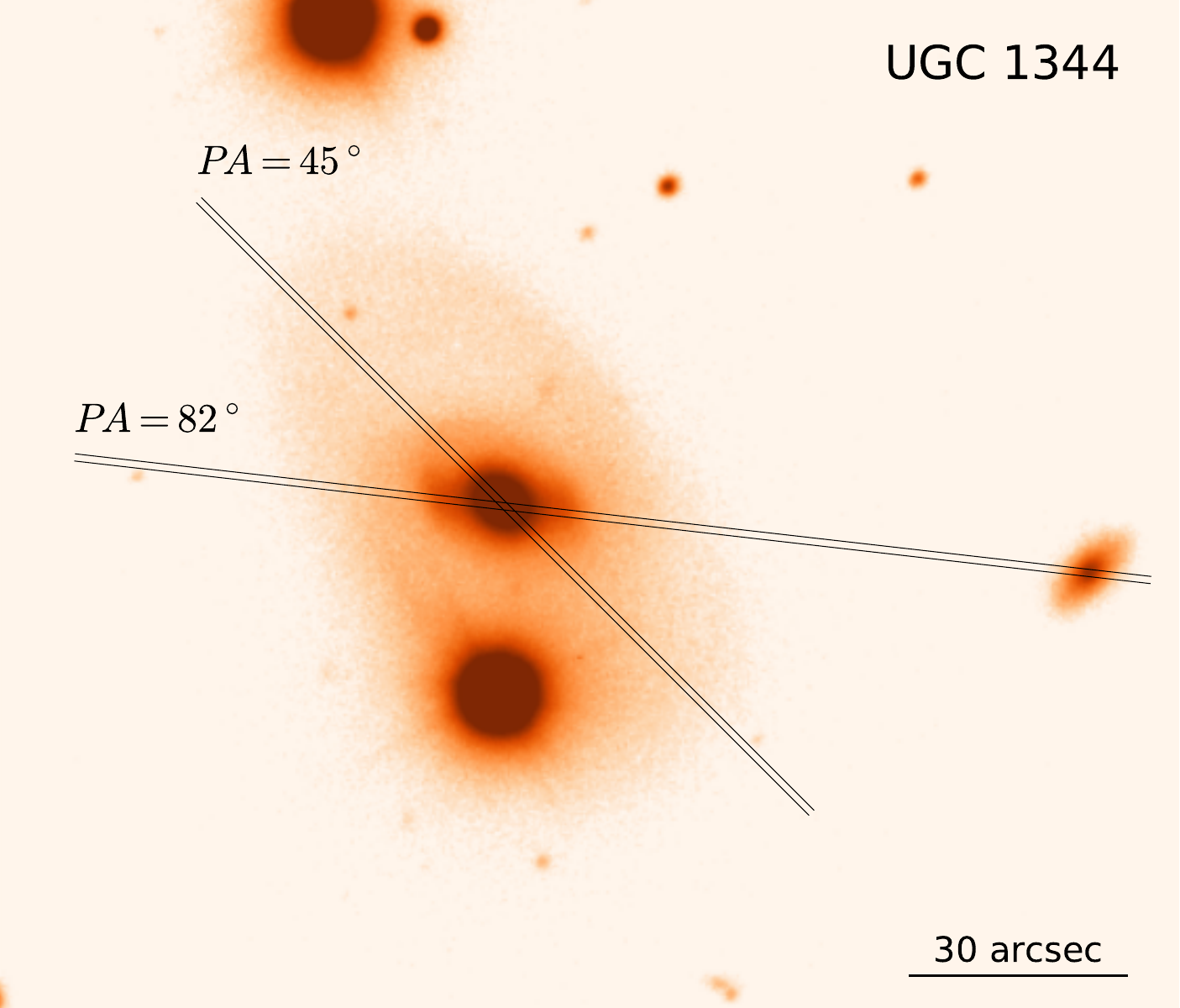}

\caption{Images of the galaxies with the overplotted positions of the slits. Left: composite SDSS $gri$-image of NGC~5347, right: $R$-band image of UGC~1344 taken with BTA.}
\label{map}
\end{figure*}

We carried out the long-slit spectral observations of NGC~5347 and 
UGC~1344 with the SCORPIO focal reducer (see \citealt{AfanasievMoiseev2005}) at the Russian 6-m BTA telescope in  2015 and 2016. In Table \ref{log} we give the position angles of the slit, the exposure times, the dates of observations and seeing. We used VPHG2300G
grism with the spectral range of 4800-5570~\AA  ~and reciprocal
dispersion  0.38\AA/px. This corresponds to spectral resolution $R \approx 2200$.
The scale along the slit is $\rm 0.36$arcsec$/px$.   Fig. \ref{map} demonstrates the slit positions for the galaxies.

The data were processed as it was described in \cite{Saburova2015}. The data reduction included the following steps: bias subtraction and truncation of overscan regions; division by normalized flat field frames; the wavelength calibration using the spectrum of  He-Ne-Ar calibration lamp; linearization and summation; the night sky subtraction; the flux calibration using the spectra of the standard spectrophotometric stars HZ2 and Feige 56. We also took into account the variation of instrumental profile of the spectrograph by analyzing the twilight sky
spectrum observed in the same observational run (for more details see, e.g. \citealt{Saburova2015}). 

After the primary reduction we fitted the spectra with high-resolution PEGASE.HR~\citep{LeBorgneetal2004} simple stellar population models (SSP) convolved with the instrumental profile. For this purpose we used the \textsc{NBursts} full spectral fitting
technique  \citep{Chilingarian2007a, Chilingarian2007b}. The advantage of this technique is that it allows to fit the spectrum in a pixel space.  In this method the parameters of the stellar populations are derived by nonlinear minimization
of the quadratic difference chi-square between the observed and model spectra. The parameters of SSP that we utilized are the age T~(Gyr) and metallicity [$Z/H$]~(dex) of stellar population. The line-of-sight velocity distribution (LOSVD) of stars were parametrized by Gauss-Hermite series (see \citealt{vanderMarel1993}).  From  the modelling we determined the luminosity-weighted stellar age and metallicity, line-of-sight velocity, velocity dispersion and Gauss-Hermite moments $h_3$ and $h_4$ which characterize the deviation of LOSVD from the Gaussian profile.

We estimated the parameter uncertainties of our stellar population model by means of Monte Carlo simulations for a hundred realizations of synthetic spectra for each spatial bin which were created by adding a random noise corresponding to the signal-to-noise ratio in the bin to the best-fitting model.

We also analyzed the emission spectra which we obtained by subtraction of model stellar spectra from the observed ones. To derive the velocity and velocity dispersion of ionized gas we fitted the emission lines by Gaussian profiles.  To increase the signal-to-noise ratio $\rm S/N$ of the spectra the adaptive binning was used in the fitting. We specified $\rm S/N=15$ for all slices besides that with $\rm PA=30\degr$ of NGC~5347 for which we adopted $\rm S/N=10$.

\begin{table} 
\caption{Log of observations}\label{log}
\begin{center}
\begin{tabular}{ccccc}
\hline\hline
Galaxy&Slit PA & Date & Exposure time& Seeing \\
   &(\degr)  &  &     (s) &        (arcsec) \\
\hline
NGC~5347&103&11.05.2016&3600&3 \\
NGC~5347&120&11.05.2016&3600&2.9 \\
NGC~5347&30&11.05.2016&1866& 3 \\
UGC~1344&45&17.09.2015&13200&1.7\\
UGC~1344&82&19.09.2015&18000&2\\
\hline\hline
\end{tabular}
\end{center}
\end{table}

\subsection{The Non-parametric LOSVD Reconstruction}\label{s_nonpar}
Parallel with the parametric LOSVD reconstruction described in the previous subsection we also applied to UGC~1344 -- the galaxy with the highest signal-to-noise ratio of the spectra -- a non-parametric recovery technique based on the full spectral fitting. The technique is described in details in \cite{Katkov2013, Katkov2016}. 

The main idea of the method is that observed galaxy spectrum logarithmically rebinned in the wavelength domain can be represented by the convolution of LOSVD and a typical stellar (template) spectrum. The reconstruction technique does not require any a priori knowledge about the LOSVD shape and searches for the solution of the convolution problem as a linear inverse ill-conditioned problem by using a smoothing regularization. We used the SSP model from the \textsc{NBursts} fitting, pre-convolved with the LSF output, as a template spectrum. In case of contribution to the integrated spectrum of very different stellar populations the technique should be applied with caution. Since the method fits observed spectrum by only one given template the mismatch template problem can lead to biased LOSVD recovery. For UGC~1344 we show below that stellar population parameters of two components are similar hence the non-parametric LOSVD recovery is not affected by mismatch template problem.

\begin{figure*}
\centering
\includegraphics[width=0.45\textwidth]{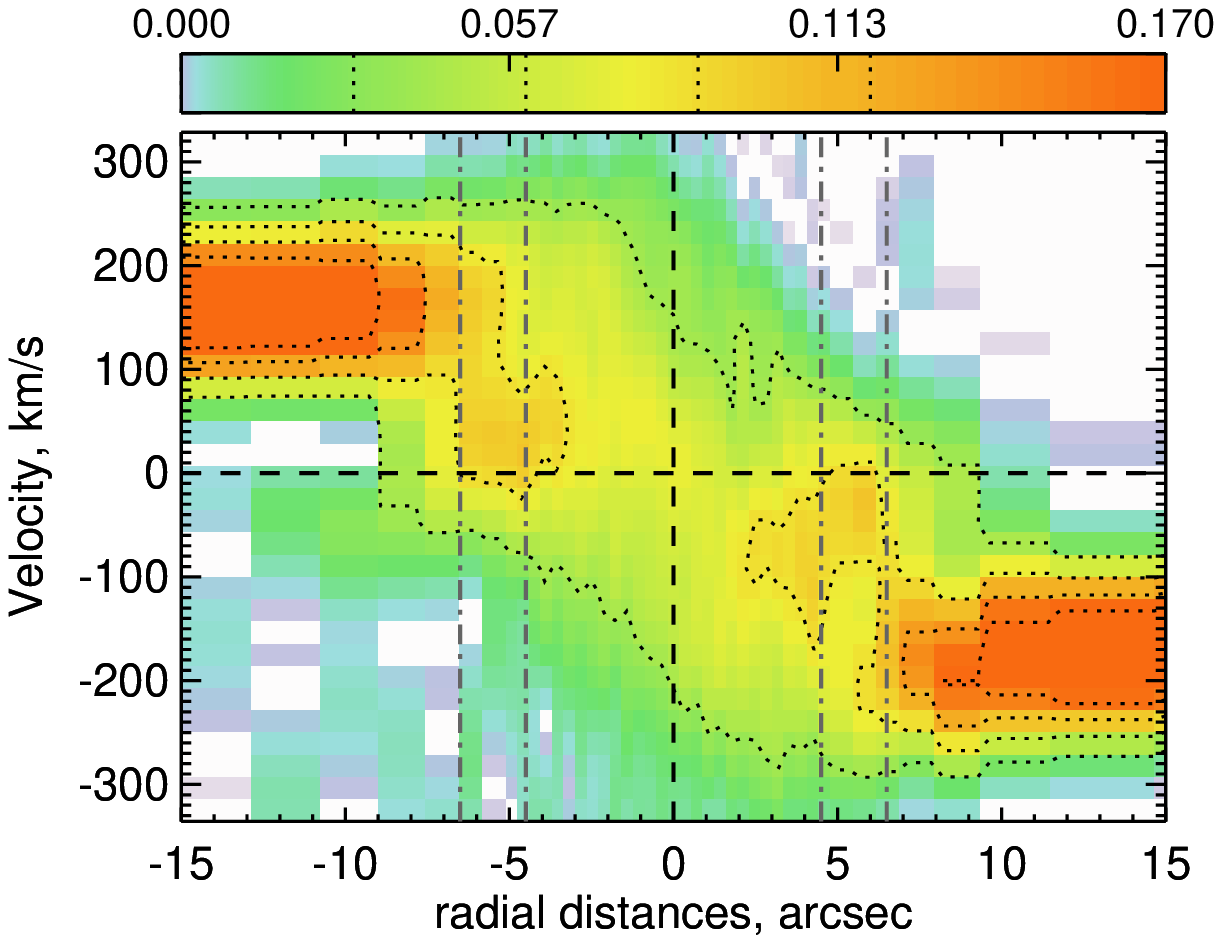}
\includegraphics[width=0.45\textwidth]{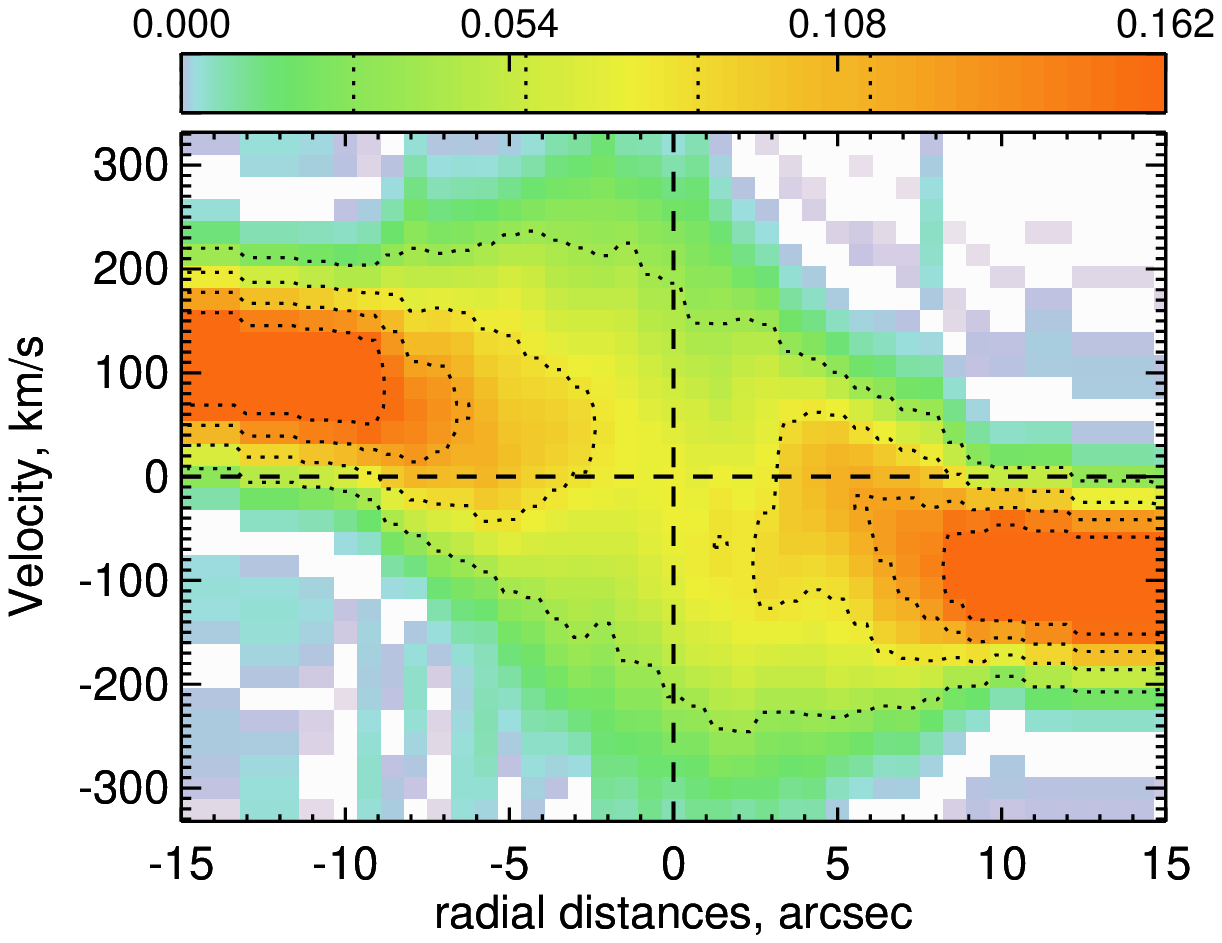}
\caption{The stellar position-velocity diagrams of UGC~1344 obtained by non-parametric LOSVD reconstruction method for $\rm PA=45\degr$ (left), $\rm PA=82\degr$ (right). Dash-and-doted lines show the transitional region where kinematically decoupled component is clearly seen. These regions were used for the non-parametric analysis as well as for the spectral decomposition. The LOSVD is normalized for given radial position.}
\label{nonpar} 
\end{figure*}

\begin{figure*}
\centering
\includegraphics[width=0.8\textwidth]{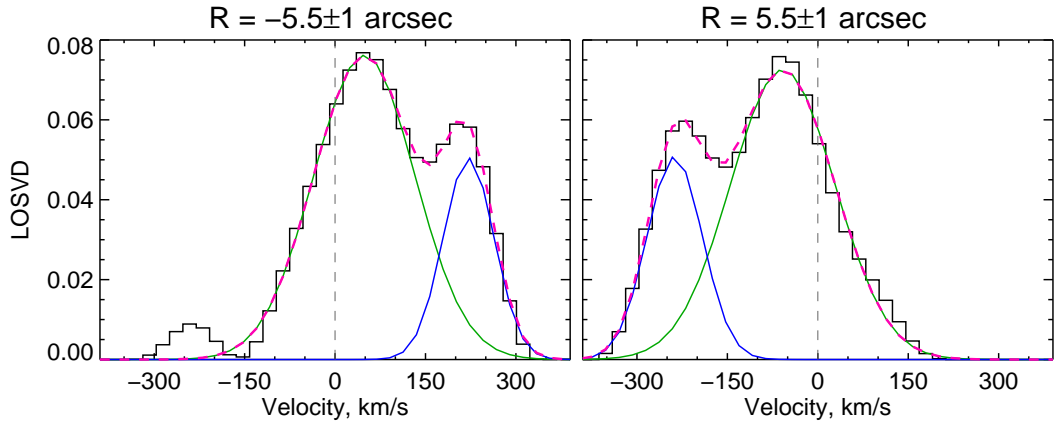} 
\caption{Decomposition of stellar LOSVD of UGC~1344 in  $\pm$5 arcsec from the nucleus where slow component is detected. Recovered stellar LOSVDs are shown by black stepped lines. Green and blue lines demonstrate LOSVD decomposition into two Gaussian components (main component and kinematically decoupled one). Pink line marks the resulting model.}
\label{nonpar2} 
\end{figure*}

\begin{figure*}
\centering
\includegraphics[width=0.8\textwidth]{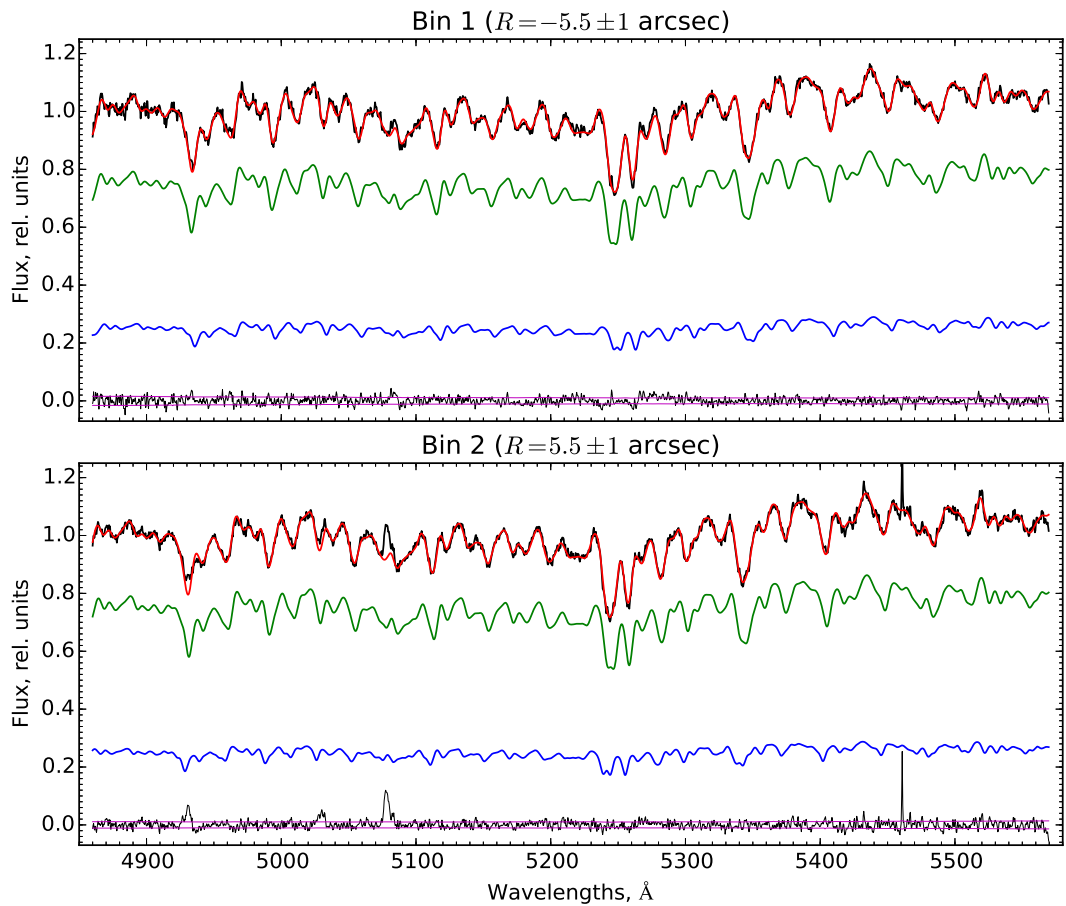} 
\caption{Spectral decomposition in two bins at $\pm$5.5 arcsec from the nucleus of UGC 1344 where kinematically decoupled component is detected. The red line corresponds to the best-fitting model overplotted on top of the observed spectrum (black line). Green and blue lines show the different components. Residuals are shown on bottom of each panel. Magenta line presents noise level of observed spectra.}

\label{spec_nonpar} 
\end{figure*}

The stellar position-velocity diagram for 
UGC~1344 obtained by this  technique for $\rm PA=45\degr$~(left) and $\rm PA=82\degr$~(right) is presented in  Fig. \ref{nonpar}. This method allowed us to reveal the asymmetry of stellar LOSVD of UGC~1344 for spectral slice  $\rm PA=45\degr$  at $r \approx\pm 5$ arcsec. We demonstrate  two-component velocity distribution at these radii in Fig. \ref{nonpar2} for two bins at  the right and left sides from the centre of the galaxy. The narrow range of the radii for the two bins corresponds to the transitional region between the slower component dominating in all central area and the faster rotating component.   Green lines denote the dominating component with slower rotation and blue lines mark another faster rotating component. The resulting model is demonstrated by pink line and the observed LOSVD is shown by black histogram.  The corresponding decomposition of the spectra in right and left bins is demonstrated in Fig.~\ref{spec_nonpar}. Red and black lines show the model and the observed spectra respectively, the two components are denoted by blue and green similarly to the previous figure.  The properties of stellar population of these components are given in Table \ref{tbl_spec_decomposition}.
 From Table \ref{tbl_spec_decomposition} one can see that the dominating component has slower rotation and higher velocity dispersion. Its age and metallicity are close to that of the second component. Thus, we can propose that the slow decoupled component has the internal origin and most probably was formed by dynamical evolution of the bar and could be the x-shaped bulge (for more details see Discussion). The fast rotating component apparently represents the main disc of the galaxy.

Note that the slowly rotating component is clearly seen  only in the cut running along the major axis. For the cut along the bar at $\rm PA=82\degr$ it shows itself very marginally. It may be explained  by the lower LOS component of rotational velocity of the bar, which makes it difficult to notice the spectral input of slowly rotating second component parallel with the fact that the slow component can have lower surface brightness in comparison to the bar.  When the slit is oriented along the bar the latter dominates the dynamics in this region making it difficult to reveal the faint detail of velocity distribution.

\begin{table*} 
\caption{Results of spectral decomposition for two positions from left and right sides from the centre of UGC 1344: the light fraction of the component, the LOS stellar velocity, the stellar velocity dispersion, SSP-age and SSP-metallicity of stars. Component weights  are the same for both bins hence only one column is shown in the table.}\label{tbl_spec_decomposition}
\begin{center}
\footnotesize
\begin{tabular}{cccccc c cccc}
\hline\hline
&&\multicolumn{4}{c}{$R=-5.5 \pm 1$ arcsec} && \multicolumn{4}{c|}{$R=5.5 \pm1 $ arcsec}\\
\cline{3-6}  \cline{8-11}
& $w_{1,2}$ & $v_1$,~\kmps & $\sigma_1$,~\kmps  & $T_{SSP,1}$, Gyr & $Z_{SSP,1}$, dex && $v_2$,~\kmps & $\sigma_2$,~\kmps  & $T_{SSP,2}$, Gyr & $Z_{SSP,2}$, dex \\
 
\hline

Cmp.1 &  0.75 $\pm$  0.02 & 4454.1 $\pm$   16.5 & 114.2 $\pm$  11.6 &  7.63 $\pm$  0.32 & -0.17 $\pm$  0.02 && 4340.7 $\pm$   15.3 & 111.2 $\pm$   8.4 &  8.21 $\pm$  0.57 & -0.18 $\pm$  0.02 \\
Cmp.2 &  0.25 $\pm$  0.02 & 4609.5 $\pm$   48.9 &  91.0 $\pm$  20.7 &  6.98 $\pm$  0.65 & -0.13 $\pm$  0.02 && 4166.4 $\pm$   45.5 &  80.4 $\pm$  29.2 &  9.18 $\pm$  0.83 & -0.25 $\pm$  0.04 \\

\hline\hline
\end{tabular}
\end{center}
\end{table*}

\subsection{Kinematical profiles}
\begin{figure*}
\centering
\includegraphics[clip,trim={0.0cm 0 0.18cm 0},height=0.81\textheight]{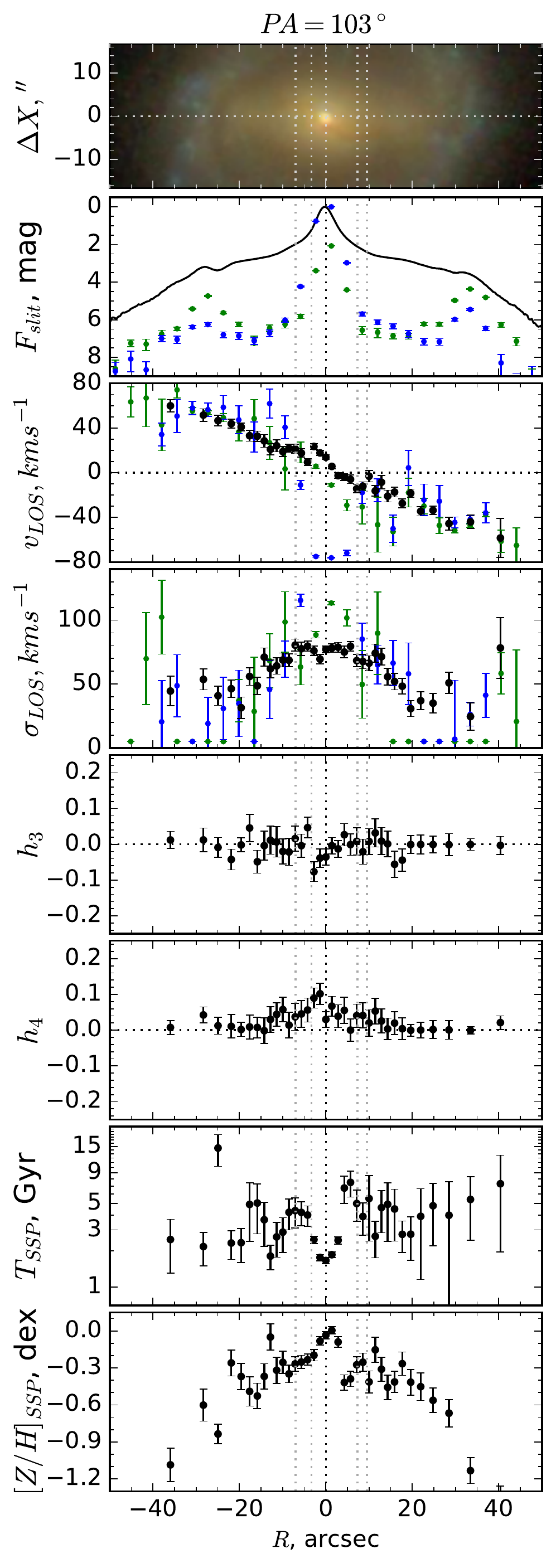}
\includegraphics[clip,trim={2.0cm 0 0.18cm 0},height=0.81\textheight]{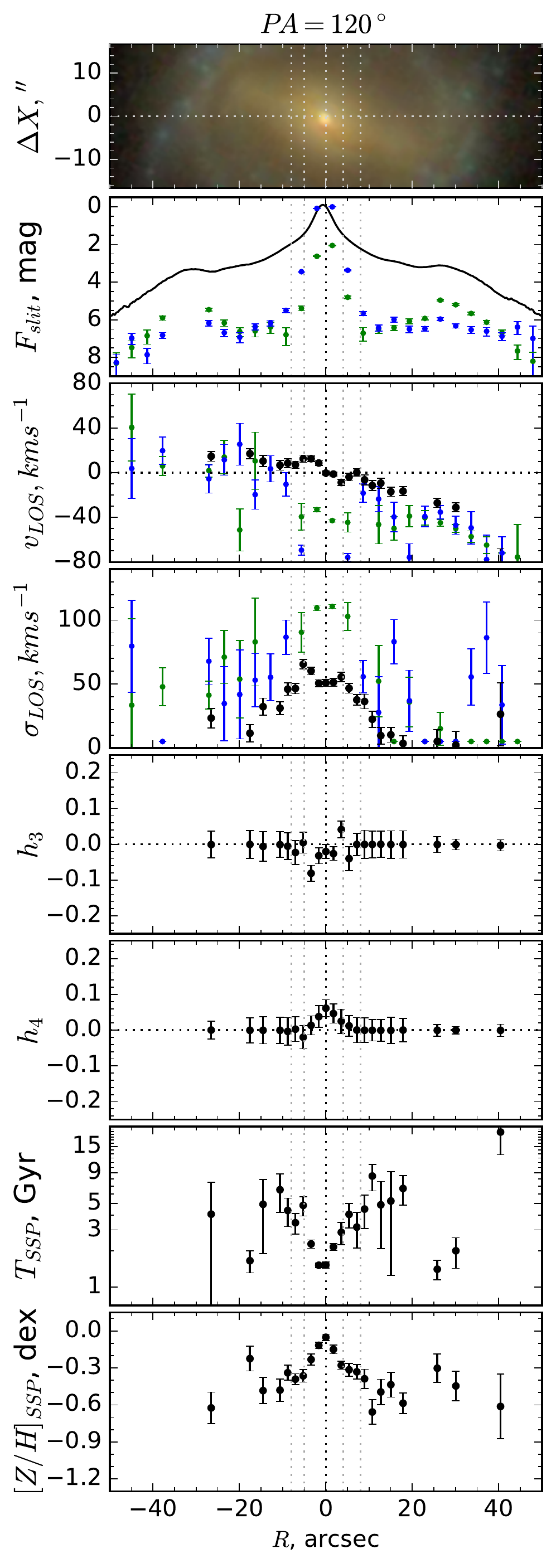}
\includegraphics[clip,trim={2.0cm 0 0.25cm 0},height=0.81\textheight]{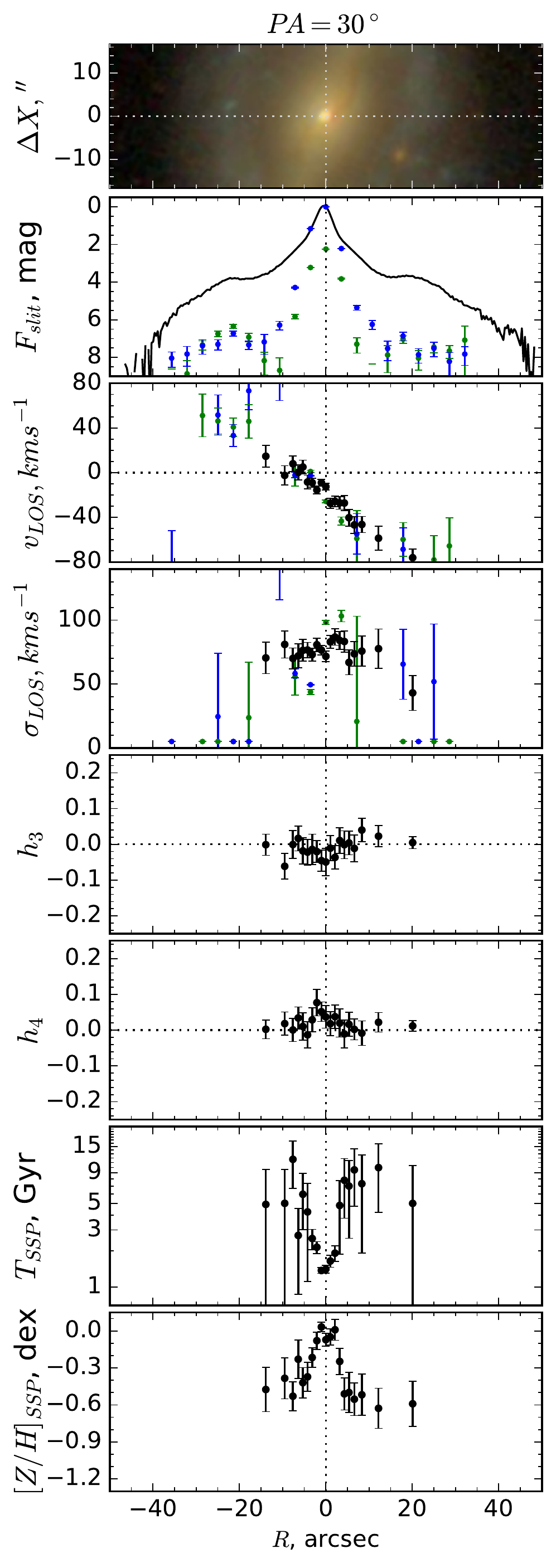}
\caption{From left to right: the radial profiles of the properties of stars and ionized gas of NGC~5347 obtained for $\rm PA=103\degr$, $\rm PA=120\degr$ and $\rm PA=30\degr$. The top panels show the reference composite {\it g,r,i} images from SDSS. Next panels correspond to the radial profiles of the total flux (black line) and flux in $\rm H \beta$  (green dots) and [OIII] (blue dots) lines. Two panels below demonstrate the profiles of the line-of-sight velocities and velocity dispersion. Black and coloured symbols correspond to stars and ionized gas respectively. Zero-point for the velocity is 2410~~\kmps. Vertical  dotted lines mark the positions of the centre of the galaxy and radial distances of the noticeable fractures  of profiles of LOS velocity. Next two panels give the radial profiles of Gauss-Hermite moments $h_3$ and $h_4$ of stars. The bottom two panels correspond to the radial distributions of age and metallicity of stars. }
	\label{profiles_n5347}
\end{figure*}

\begin{figure*} 
\includegraphics[clip,trim={0.0cm 0 0.2cm 0},height=0.81\textheight]{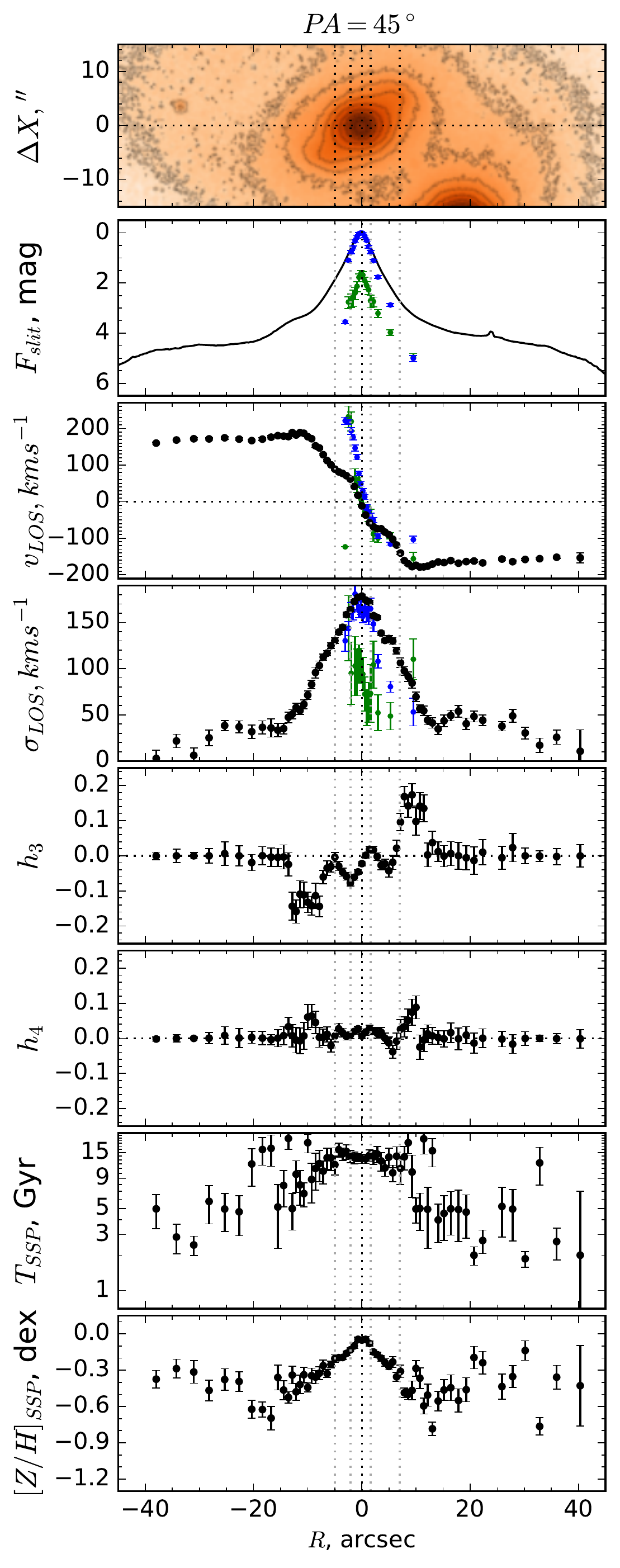}
\includegraphics[clip,trim={2.8cm 0 0.3cm 0},height=0.81\textheight]{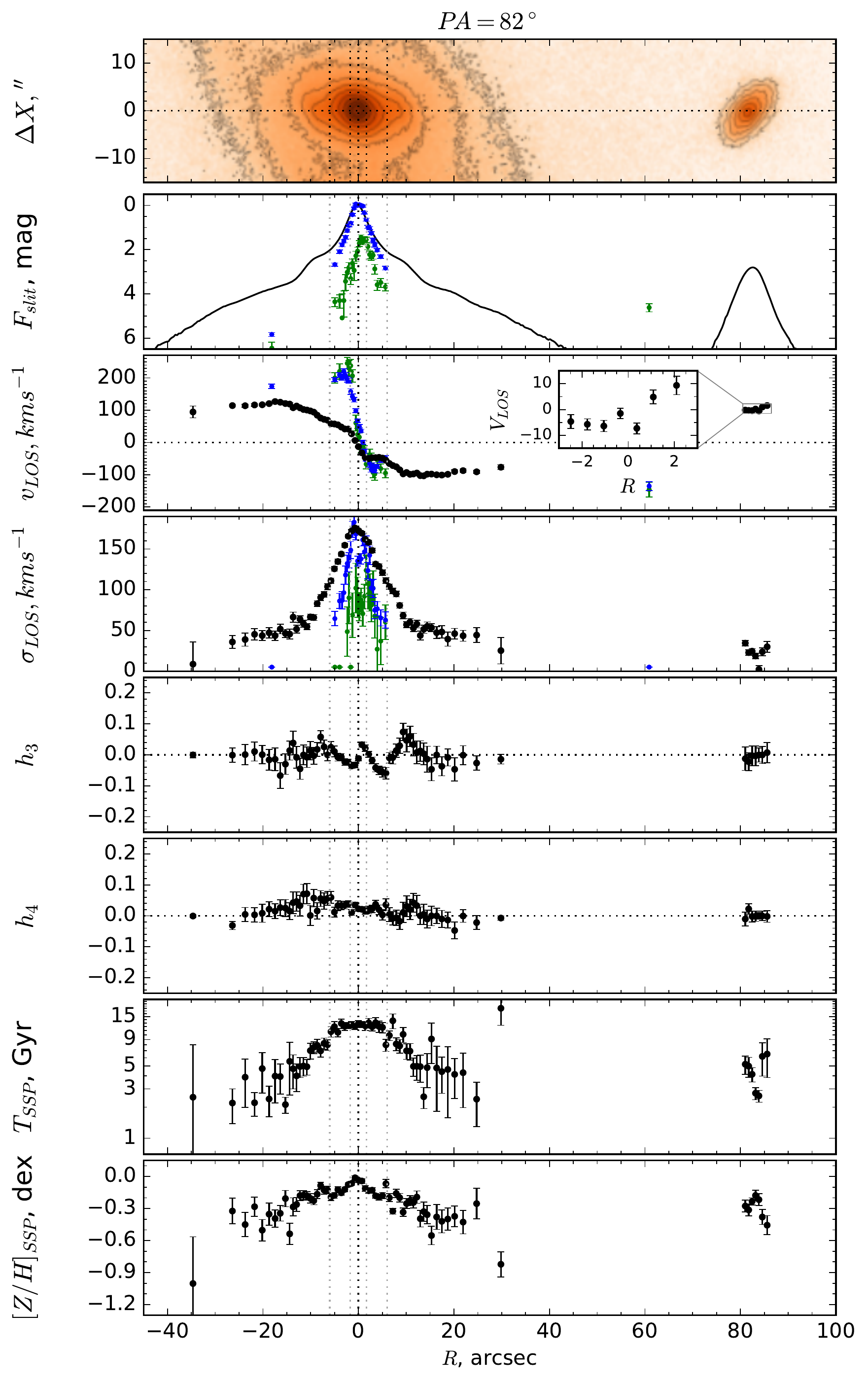}
\caption{From left to right: the radial profiles of properties of stars and ionized gas of UGC~1344 obtained for $\rm PA=45\degr$ and $\rm PA=82\degr$. The top panel shows the reference images from SCORPIO. Next panels correspond to the radial profiles of the total flux (black line) and flux in $\rm H \beta$  (green dots) and [OIII] (blue dots) lines. Two panels below demonstrate the profiles of the line-of-sight velocities and velocity dispersion. Black and coloured symbols correspond to stars and ionized gas respectively. Zero-point for the velocity is 4390~\kmps. Vertical dotted lines mark the positions of the centre of the galaxy and radial distances of the noticeable fractures  of profiles of LOS velocity. The insertion in the velocity profile for PA=82\degr shows the zoomed in velocity profile for the companion 2MASX J01522797+3629533. Next two panels give the radial profiles of Gauss-Hermite moments $h_3$ and $h_4$ of stars. The bottom two panels correspond to the radial distributions of age and metallicity of stars.}
	\label{profiles_u1344}
\end{figure*}

In Figs. \ref{profiles_n5347} and \ref{profiles_u1344} (the third and fourth panels from top) we show the radial profiles of line-of-sight velocity and velocity dispersion of stars and ionized gas for different spectral slices of NGC~5347 and UGC~1344 respectively obtained by parametric LOSVD reconstruction. In these figures we also give the reference images from SDSS and SCOPRIO (top panels) and the radial profiles of fluxes of  continuum and emission lines  (second panels from top). 

It is worth noting the unusual behavior of the emission gas velocities in  the central regions of the galaxies. In both galaxies the gas possesses  high velocity dispersion here, which is equal or exceeds the stellar one.  In NGC~5347  the LOS gas velocities are shifted at 40- 80 ~\kmps towards the observer with respect to the stellar ones along the slits PA = 120,103\degr, which is natural to connect with the AGN activity in this galaxy. The ionized gas velocity profile of UGC~1344 is asymmetric and rises much steeper than that of stars in the left half of the galaxy, which is evident from both spectral slices. Moreover, the deviation of the ionized gas velocity profile is more prominent for [OIII] line, than for H$\beta$ especially for the slice with $\rm PA=45 \degr$. The flux ratio  [OIII] vs $\rm H \beta$ lines is quite high, it roughly equals to 5 in the centre, increasing to the left part of the galaxy and decreasing to the opposite side (see Fig. \ref{profiles_u1344}, second panel from top). The high flux ratio together with high velocity dispersion of ionized gas is a clear indication of  AGN in the galaxy which also could be one of the causes of the asymmetry of velocity distribution noted above. Another probable cause of the asymmetry is the gas interaction with  the contrast bar in the presence of dust absorption, which is usually observed along the bar in SB-galaxies.

As one can see from the figures, the radial profiles of stellar line-of-sight velocity of both galaxies are not smooth for the position angles close to the major axis. There are regions where the velocity gradient changes its sign. We mark  by dotted vertical lines the  areas where the steep velocity profile  becomes shallower or even begins to decrease. This behavior is typical for barred galaxies. The velocity turn-over was already observed in eighties by \citet{PetersonHuntley1980}  in barred galaxy NGC1300.  It was also found in edge-on systems with boxy or peanut-shaped bulges by \cite{ChungBureau2004} and predicted by simulations of \citet{BureauAthanassoula2005}. The double-hump in the velocity profile was found in 60 percent of barred galaxies in \cite{Seideletal2015} and was considered as a hint towards the existence of inner discs or rings in barred galaxies. 

The stellar velocity dispersion profiles demonstrate a different behavior for NGC~5347 and UGC~1344. For NGC~5347 they show central minima for all  spectral slices. Similar minima were found previously for many barred galaxies (see e.g., \citealt{Bettoni1997}). The presence of the minimum of the stellar velocity dispersion can give evidences in favor of the inner disc. At the same time for UGC~1344 we observe a peak of the velocity dispersion and the  flattening of the decrease for both sides from the centre. Evidently this galaxy possesses more concentrated and massive stellar bulge.

We did not find any significant changes of the velocity and velocity dispersion in the regions of the ends of the bars.

In fifth and sixth panels from top of Figs. \ref{profiles_n5347} and \ref{profiles_u1344} we show the radial profiles of Gauss-Hermite moments $h_3$ and $h_4$ of stars for NGC~5347 and UGC~1344. These parameters characterize the deviation of LOSVD from the Gaussian profile. The value of $h_3$ is related to the skewness of the LOSVD, while $h_4$ is a measure of the kurtosis. As one can see from the profiles both $h_3$ and $h_4$ moments differ from zero in the central parts of the galaxies. The values of $h_3$ and $h_4$ are lower for NGC~5347 than that in UGC~1344. It could be because the stellar velocity dispersion of NGC~5347 is also significantly lower than that in UGC~1344 and is compatible with the resolution. The important diagnostic feature is that $h_3$ anticorrelates with the line-of-sight velocity for both galaxies in the centre. It could also evidence in favor of inner discs or rings (see, e.g. \citealt{Seideletal2015}).

\subsection{The  properties of stellar population}
Two bottom panels of Figs. \ref{profiles_n5347} and \ref{profiles_u1344} demonstrate the radial profiles of luminosity-weighted age and metallicity of stellar population for NGC~5347 and UGC~1344 correspondingly.  The age profile in NGC~5347 reveals a depression of stellar age in the centre, presenting at all spectral slices.  The age of the nuclear component of NGC~5347 is close to that of the main disc, while the bar is relatively old. This finding is in  good agreement with the conclusion made for several barred galaxies by \citet{Seideletal2016} evidencing the low or absent star formations in bars of early-type discy galaxies and more extended star formation history in the very inner region. In contrast, UGC~1344 does not demonstrate the central decrease of age, instead it has a flat inner part: stellar population is old there all over the central region. 

In both cases the ages of stellar population in bars are relatively old: $5-10$~Gyr confirming that that a bar is long-lived structure in  good agreement with the previous studies (see, e.g. \citealt{ShenSellwood2004}, \citealt{GadottideSouza2006}; \citealt{Perez2007}).
We also observe the flattening of the stellar metallicity gradient in the regions of bars where  the velocity gradient changes -- the effect possibly caused by the stellar radial migration. \cite{Seideletal2016} also found the metallicity flattening, however only for bar major axis.

Interestingly, while the metallicity profiles of UGC~1344 and NGC~5347 show similar nearly solar metallicities in the centres, the situation is different at the outskirts of galaxies: the stellar metallicity profiles of NGC~5347 demonstrate a decrease in the outer radii especially for PA=103\degr not presented in UGC~1344. Note that the $B-V$ and $V-R$ azimuthally averaged radial profiles do not show a sharp decrease in the outer region of NGC~5347 (see \citealt{Saburovaetal2011}). It could indicate that the colour indices reflect a current rate of star formation, not the stellar underabundance.

\section{Discussion}\label{Discussion}
\subsection{On the mass-to-light ratios}
As we noted in Section \ref{intro} the considered galaxies were chosen as the candidates of systems with low mass-to-light ratios of stellar population. The new kinematic data allowed us to verify this peculiarity.

\subsubsection{UGC~1344}
For UGC~1344 all previously available estimates of the rotation velocity were based on the measurements of the H{\sc i} linewidth. Different sources give the following values corrected for the same inclination taken from the Hyperleda database (see Table \ref{properties}): $v=40$~\kmps (Hyperleda);  $v=56$~\kmps  (RC3, \citealt{rc3}); $v=59$~\kmps (\citealt{Wilkerson1980}). All these estimates correspond to low dynamical mass-to-light ratio within the optical radius: $M/L_B < 0.34$, which is more than 6 times lower than expected from the observed red colour $(B-V)_0$ and stellar model relations for scaled Salpeter IMF \citep{bdj}.  From the radial profiles of velocity obtained in the current article for UGC~1344 (see Fig. \ref{profiles_u1344}) it follows that the rotational velocity amplitude corrected for inclination is at least 3 times higher than that determined from H{\sc i} linewidth: $v=196$~\kmps. It corresponds to the dynamical mass-to-light ratio $M/L_B = 3.75$, which is quite normal for spiral galaxies containing  a comparable fractions of light and dark matter within the optical borders. 

One of possible explanations of the differences in the rotation velocity estimates could be a  confusion of UGC~1344 with the small companion seen at the distance of about $1.4$ arcmin from UGC~1344 with the systemic velocity of  4490~\kmps (the FWHM of the beam is $\sim 3.2$ arcmin in \citealt{Wilkerson1980}). This companion was crossed by the slit with $\rm PA=82\degr$. In Fig. \ref{profiles_u1344} we also demonstrate its radial profiles of line-of-sight stellar velocity and velocity dispersion. From this figure one can see that the slit is not oriented along the major axis of the companion. From the image of this galaxy it follows that the angle between the slit and major axis is $\phi\approx 45 \degr$ and inclination of the companion is $i\approx 70 \degr$. Assuming these values one can calculate roughly the rotational velocity amplitude of this galaxy: 
\begin{equation}
\label{formula_v} \Oo v(R) = \frac{v_r(R)\sqrt{(\sec^2(i)-\tan^2(i)\cos^2(\phi))}}{\sin(i) \cos(\phi)}\,,
\end{equation}
where $v_r\approx10$~\kmps is the total range of line-of-sight velocity of this galaxy corrected for the systemic velocity and divided by~$2$.  Then the rough estimate of the rotation velocity of the companion is $v\approx 44$~\kmps\footnote{the asymmetric drift correction will increase this value, but much less than by 3 times.} which is close to that found from H{\sc i} linewidth. However this explanation looks not too realistic since the companion does not seem to be a gas-rich galaxy (we do not see bright emission lines in its spectrum).

Another possible explanation of the difference between the obtained velocity of UGC~1344 and that based on H{\sc i} linewidth is the concentration of the neutral hydrogen in the inner part of the galaxy where the rotational amplitude is low, so the gas traces only the central part of the rotational curve. However H{\sc i} data with good resolution and sensitivity are needed to verify this hypothesis. 

\subsubsection{NGC~5347}
For NGC~5347 the situation is different. This galaxy is more studied in comparison to UGC~1344. The rotational amplitudes were available both from long-slit optical observations (\citealt{Marquez2004}): $v/\sin(i)=56$~\kmps, H{\sc i}  observations (\citealt{Noordermeer2005}): $v/\sin(i)\approx 77$~\kmps (though with low resolution), and H{\sc i} linewidth measurements: $v/\sin(i)=61$~\kmps (RC3, \citealt{rc3}). Here we used the inclination taken from Hyperleda, which is close to the value found in \cite{Saburovaetal2011} for $R=20$ arcsec.  The ionized gas rotational amplitude measurements of \cite{Marquez2004} are in satisfactorily agreement with the current article, although we failed to detect counter rotation of ionized gas found in central $5$ arcsec by \cite{Marquez2004} for $\rm PA=40\degr$ and $\rm PA=130\degr$. However, our ionized gas data are uncertain in the inner part of the velocity profile due to the presence of AGN that we did not take into account in the analysis. By the same reason we give  preferences to  stellar velocities. However, as far as the rotational amplitude is rather low, the asymmetric drift correction may be very critical for this system. 
We calculated a rough correction of  stellar rotation velocity for asymmetric drift in the epicyclical approximation for exponential stellar disc using following equation based on that from \cite{BinneyTremaine2008}:
\begin{equation}
\label{formula_drift}\Oo v_c^2 = v_r^2 + \sigma_r^2  \left(0.5 \frac{d\ln(v_r)}{d\ln(R)} -0.5 +\frac{R}{r_d} -  \frac{d \ln (\sigma_r^2)}{d\ln(R)}\right)\,,
\end{equation}
where $v_c$ is the circular velocity, $v_r$ is the observed rotation velocity of stars, $\sigma_r$ is radial velocity dispersion of stars, $r_d$ is the exponential scalelength of disc taken from \cite{Saburovaetal2011}. 
We estimated
radial velocity dispersion from the observed line-of-sight stellar velocity
dispersion $\sigma_{obs}$ at $\rm PA=103\degr$,  taking into account the expected links
between the dispersion along the radial, azimuthal and vertical directions:
\begin{equation}
\begin{aligned}
\label{formula_sigma_obs}\Oo \sigma_{\rm obs}^2(R) = \sigma_z^2 \cdot \cos^2 (i)+ \sigma_{\phi}^2\cdot \sin^2 (i)\cdot \cos^2(\alpha)+\\ +\sigma_{r}\sin^2(i)\cdot \sin^2(\alpha)\,,
\end{aligned}
\end{equation}
where $\alpha$  is the angle between the  slit and the major axis, $\sigma_z$, $\sigma_{\phi}$, $\sigma_{r}$ are the dispersions along the vertical, azimuthal and radial directions, projected upon the plane of the galaxy respectively.

To solve the equation~\ref{formula_sigma_obs} we used two additional conditions: $ \Oo \sigma_{r} = 2\Omega \sigma_{\phi} /\varkappa $ (Lindblad formula for the epicyclic approximation), where $\varkappa$ is the epicyclic frequency: $\Oo \varkappa(R)=\frac{2v(R)}{R}\sqrt{\frac{1}{2}+\frac{R}{2v(R)} \frac{\partial v(R)}{\partial R}}$ and $\Oo \sigma_z =k \sigma_{r} $. The coefficient $k$  was taken to be 0.6 in accordance with direct measurements of other galaxies, which show that it could lie in the range $0.5-0.8$~(see e.g.
\citealt{Shapiro2003}).

The asymmetric drift correction may increase the velocity by roughly  a factor of  two. The corrected value of the maximal velocity is $v_c\approx100$~\kmps which is higher than the rotational amplitude related to the disc  \citep{Saburovaetal2011} and corresponds to quite normal dynamical mass-to-light ratio $\rm M/L_R=2.3$ which is higher than that expected from the observed colour index for scaled-Salpeter IMF for the disc: $\rm (M/L_R)_d=1.3$ and for the galaxy as a whole: $\rm (M/L_R)_{tot}=1.79$ evidently due to the presence of dark halo. Thus, our new kinematical data do not confirm the unusually low mass-to-light ratio and hence the peculiar stellar IMF of this galaxy.

Of course, for both galaxies the dynamical mass estimates depend on the adopted inclination angle. However, both galaxies have moderate inclinations (they are not seen close to face-on), and the expected uncertainties of the inclination are significantly lower than the difference between the new and old measurements of the circular velocity. 
\subsection{On the central components. A comparison with the $N$-body models.}
Despite the fact that we did not confirm the peculiarity of stellar population in the considered galaxies they are certainly worth studying. These galaxies possess some remarkable features in both kinematical profiles and the profiles of the properties of stellar population. For both galaxies we obtained the so-called double-humped rotation curves -- the rise of the velocity is followed by the local decrease or flat spot after which the increase continues. The profiles of $h_3$ moment related to the skewness of the LOSVD are not equal to zero and anticorrelate with the line-of-sight velocity in the central part for both galaxies. These two factors can be regarded as indication of the presence of inner disc or ring (\citealt{Seideletal2015}). For NGC~5347 the additional confirmation of the presence of central component could come from the prominent double-armed nuclear spiral structure found by \cite{Martini2003} based on HST $H$- and $V$-band images. However, this structure could also be related to the motions of gas due to the bar.

In order to figure out what kind of structural details could be responsible for the observed features on the kinematical profiles we compared the results of $N$-body simulations of barred galaxies to our galaxies. Numerous and significant efforts were previously done in this field (see e.g. \citealt{Combes1981};\citealt{Debattista2000}; \citealt{Athanassoula2003}; \citealt{BureauAthanassoula2005}; \citealt{Wozniak2003}; \citealt{Minchev2010}).  

\begin{figure*}
\includegraphics[width=0.5\hsize]{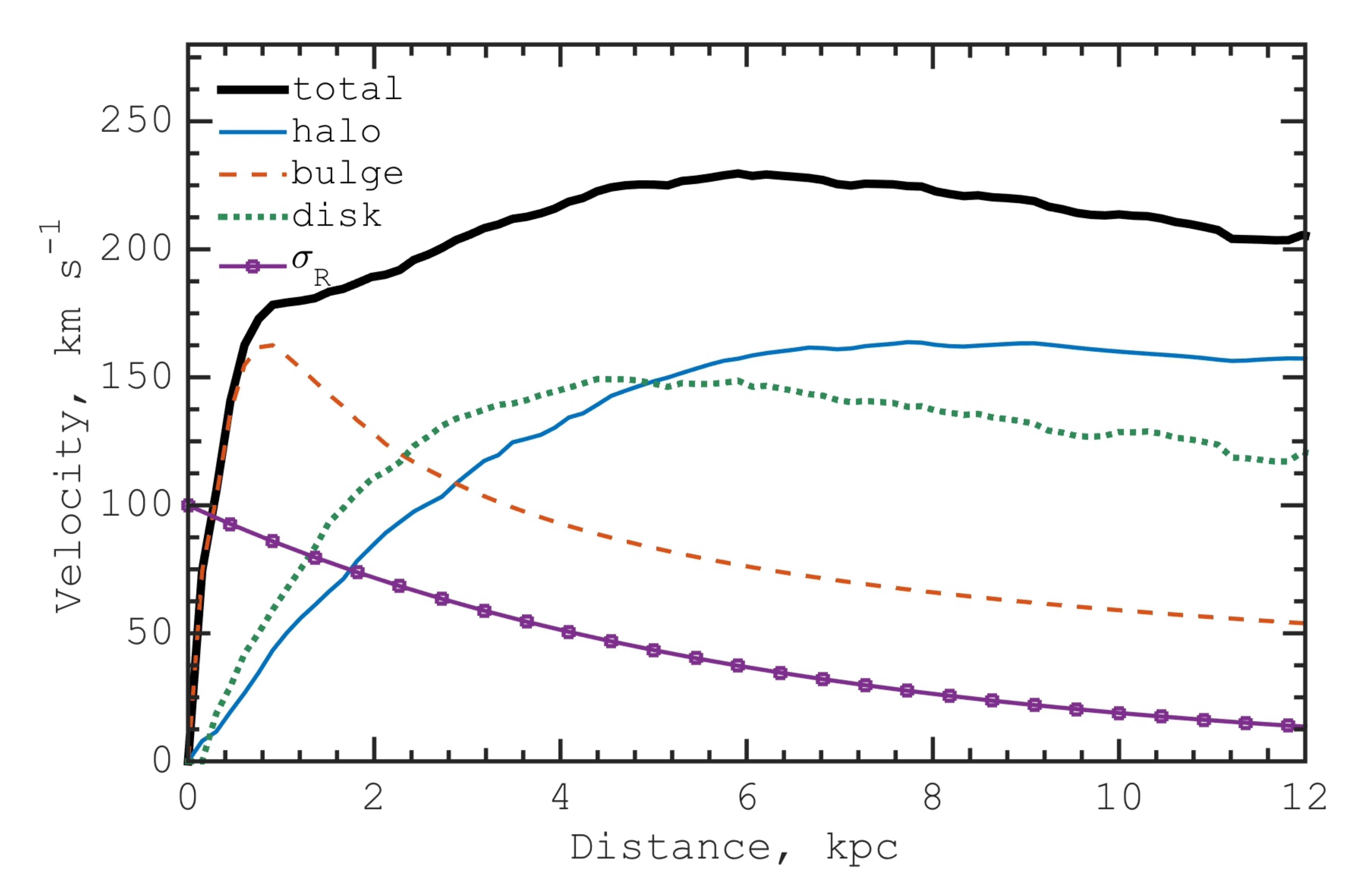}\includegraphics[width=0.5\hsize]{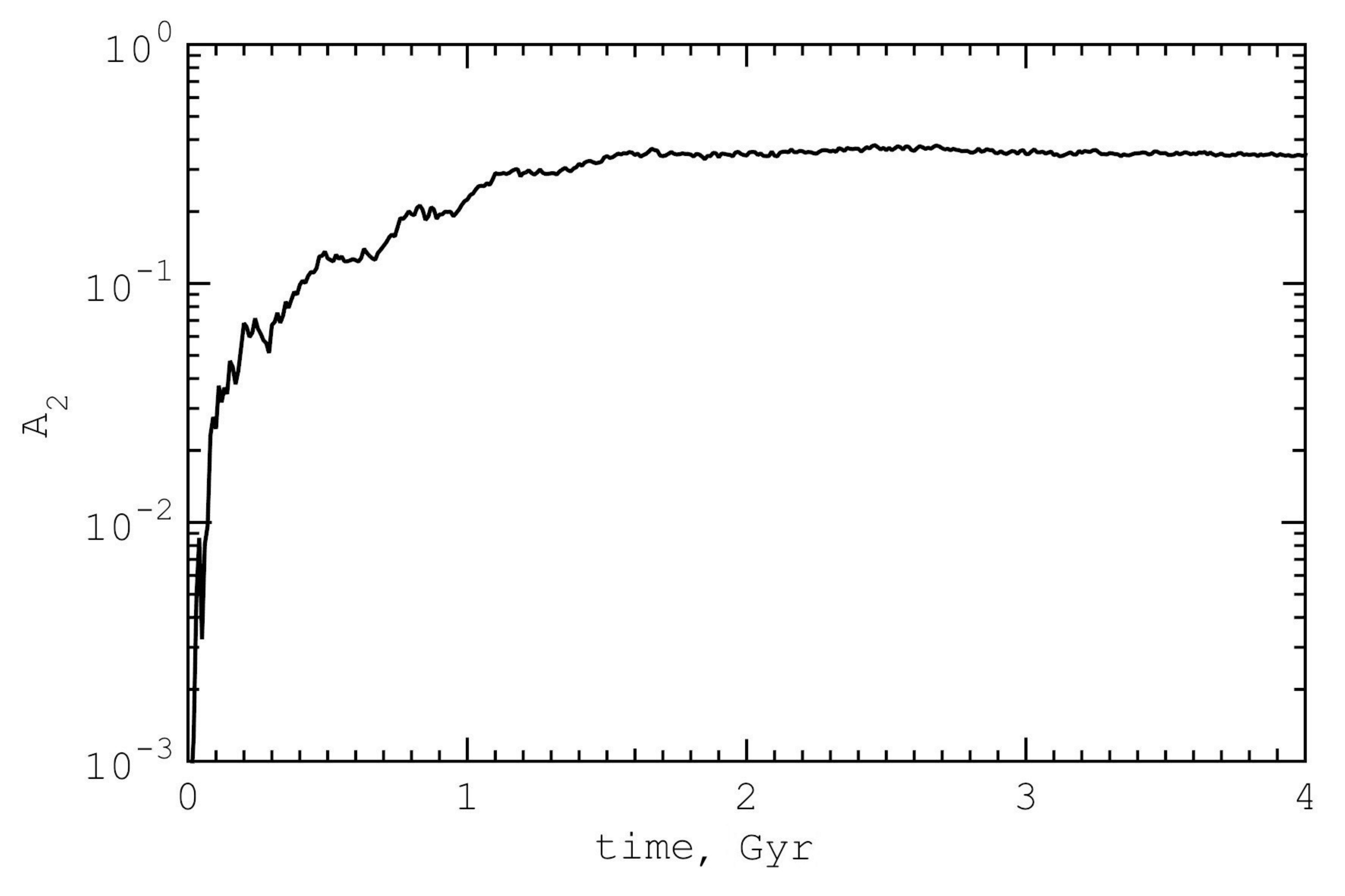}\caption{Left: rotation curve decomposition of simulated galaxy as well as the radial velocity dispersion of the stellar disk $\sigma_{\rm r}$ at the beginning of simulation. Right: strength of the bar in terms of $A_2$ as a function of time.}\label{fig::simul_1d}
\end{figure*}

Our simulations were performed with a Tree $N$-body code~\citep{2014JPhCS.510a2011K}. We set up a disc galaxy that consists of a live dark matter halo, stellar bulge and pure stellar disc. The model disc has an exponential surface density profile with total mass $3.8\times10^{10}$~\Msun and scalelength $r_d=2.5$~kpc. The vertical profile is Gaussian with a scale height $z_d = 0.3$~kpc. The dark matter halo density profile follows that of pseudo-isothermal sphere with mass $8.6\times10^{10}$~\Msun and scalelength $2$~kpc. Initially the halo is non-rotating.  We adopted bulge as a Plummer sphere with mass $6\times10^9$~\Msun and size of $1$~kpc. For the experiment described below, we generate a model containing $10^6$ disc particles, $10^6$ halo particles and $2\times10^{4}$ particles for the stellar bulge. The initial rotation curve and contributions of separate components together with the equilibrium radial velocity dispersion of stellar disc $\sigma_{r}$ are shown in Fig.~\ref{fig::simul_1d}~(left). The opening angle of TreeCode is taken to be $0.5$ and softening radius is $50$~pc.

\begin{figure*}
\includegraphics[width=1\hsize]{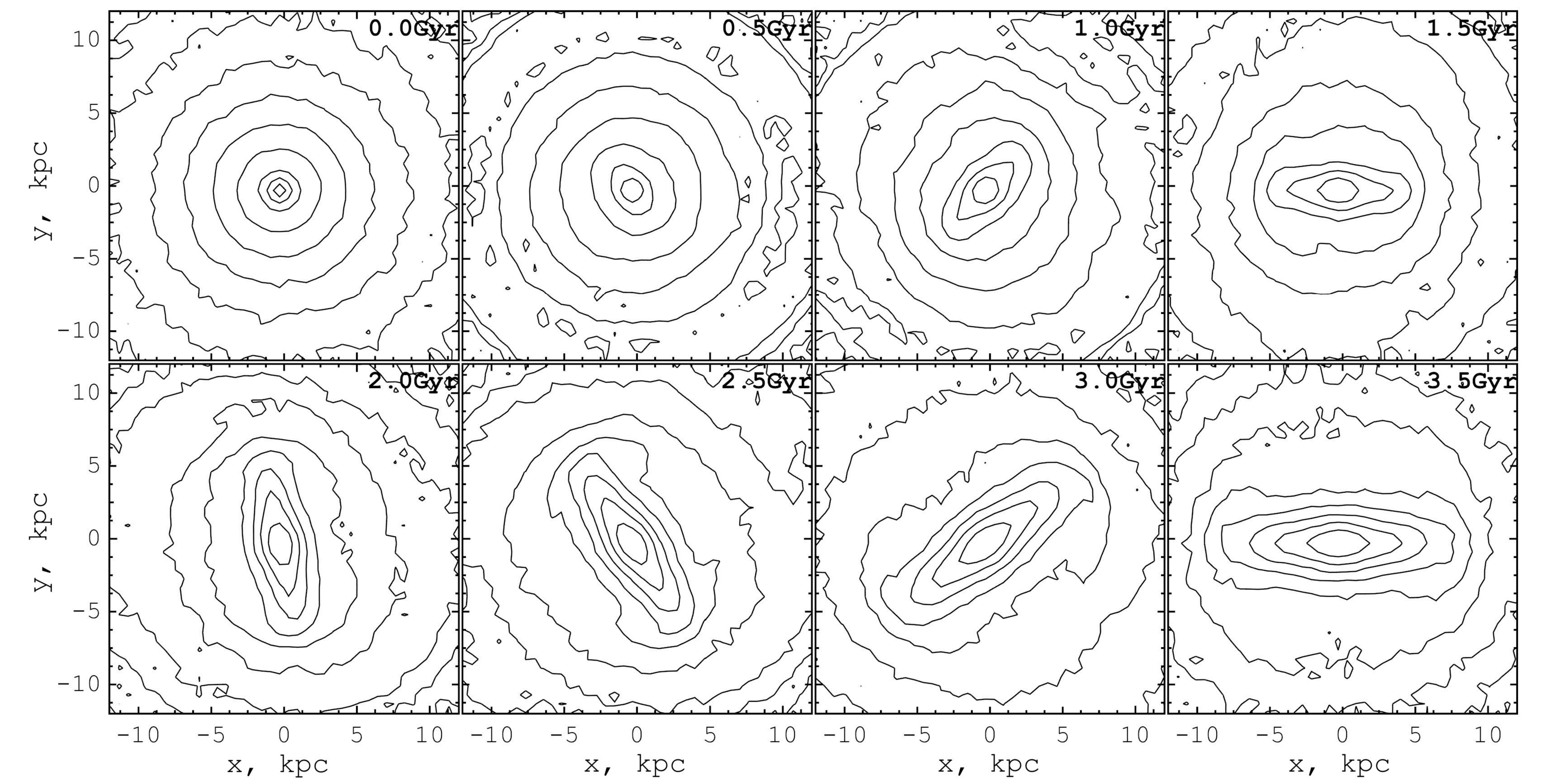}\caption{Formation and evolution of bar in the simulated galaxy.}\label{fig::evolution_bar}
\end{figure*}

The simulated galaxy evolved for $5$~Gyr. We describe the bar strength as  $m=2$ Fourier moment $A_2$. Figure~\ref{fig::simul_1d}~(right) shows that the bar strength grows since the beginning of simulation and reaches a saturation level at about $2$~Gyr. Time dependent evolution of face-on stellar density is given in Fig.~\ref{fig::evolution_bar}. Eventually the disc develops a strong bar with the size $\approx 5-7$~kpc. Our analysis of the LOS velocity distribution we address to the time $T = 2$~Gyr.

\begin{figure*}
\includegraphics[width=0.5\hsize]{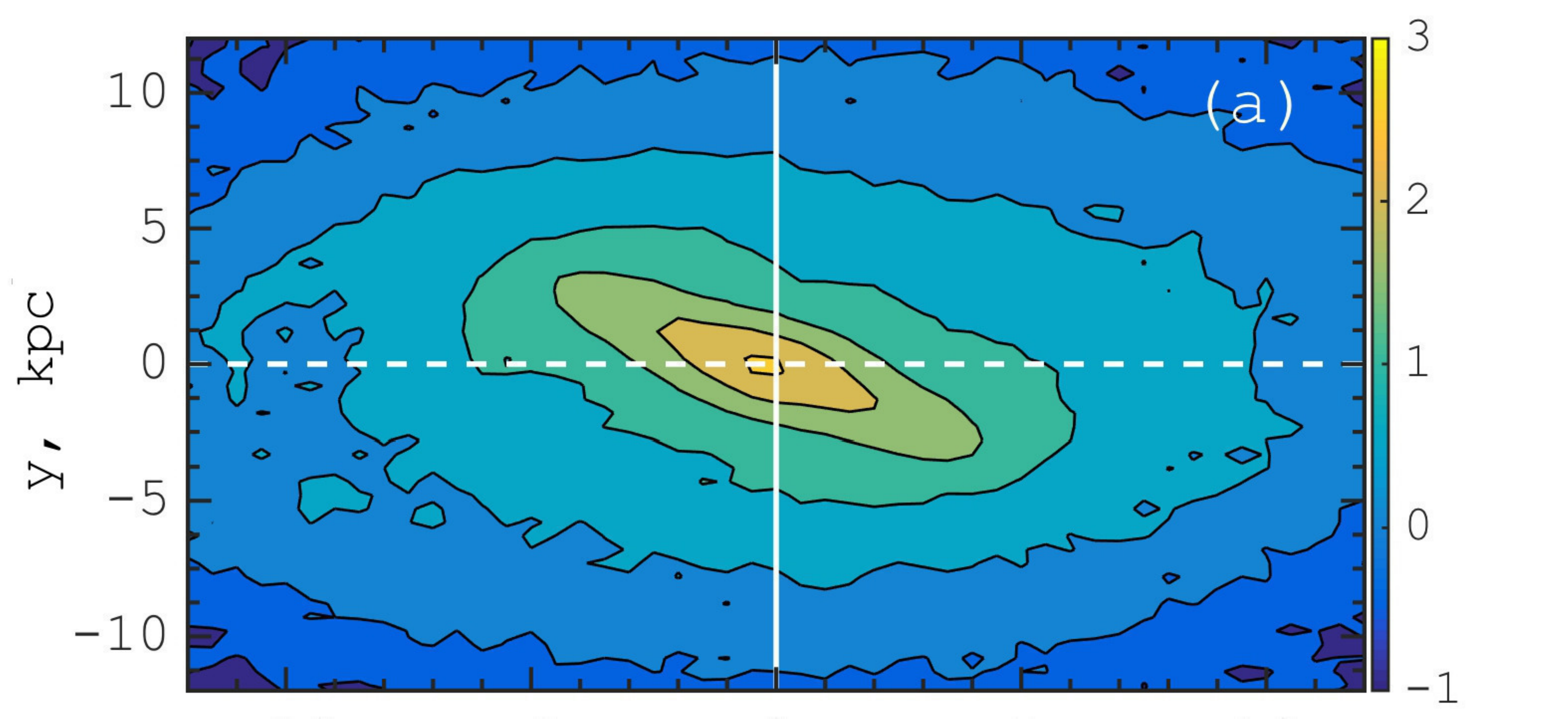}\includegraphics[width=0.5\hsize]{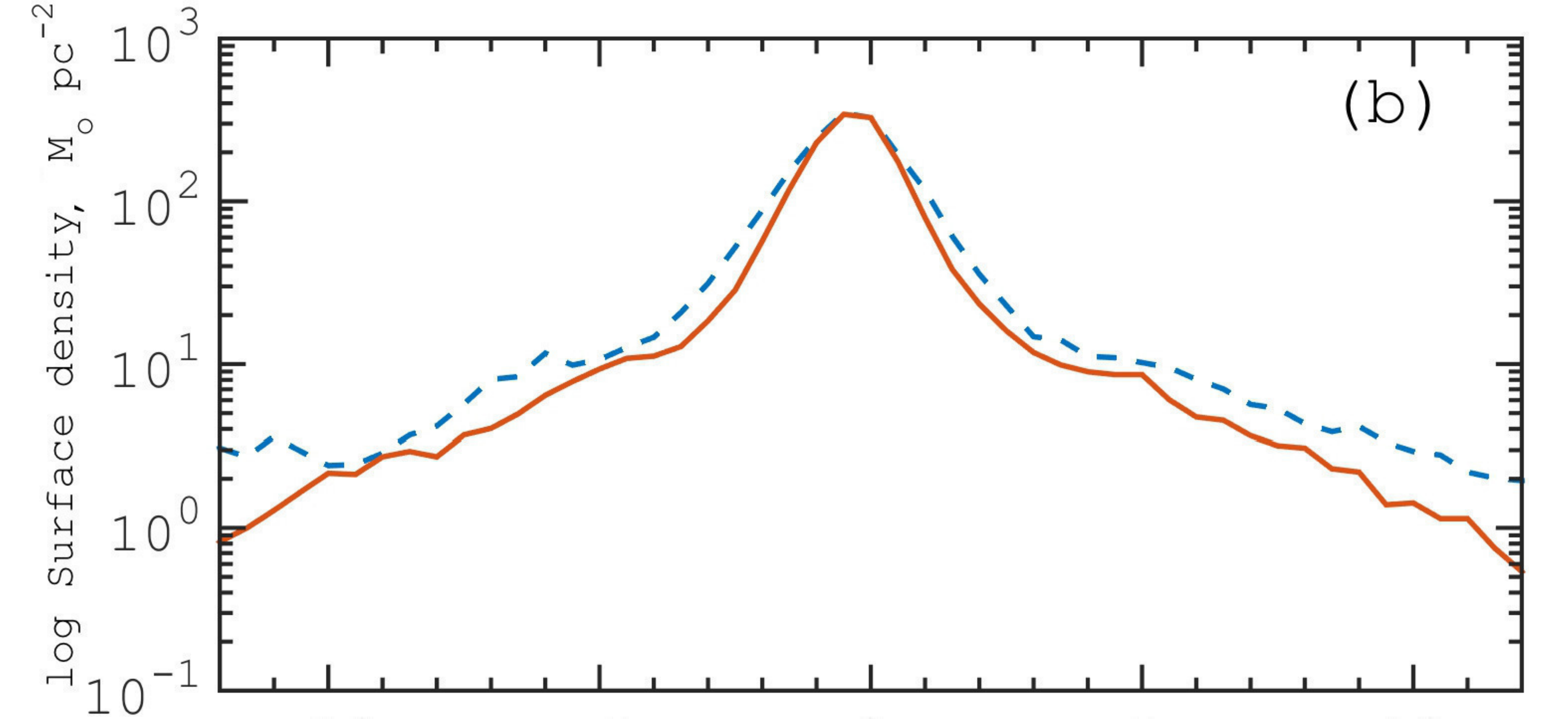}\\\includegraphics[width=0.5\hsize]{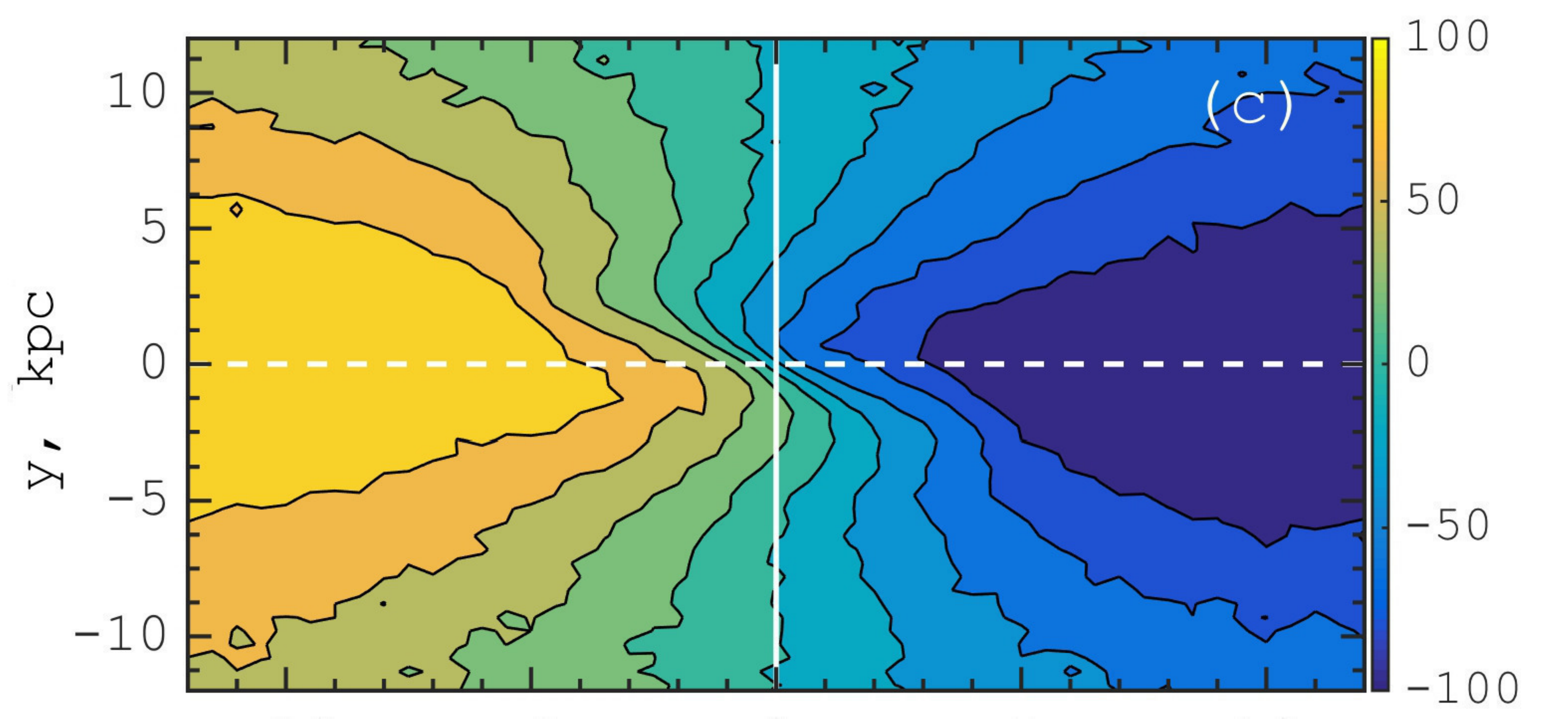}\includegraphics[width=0.5\hsize]{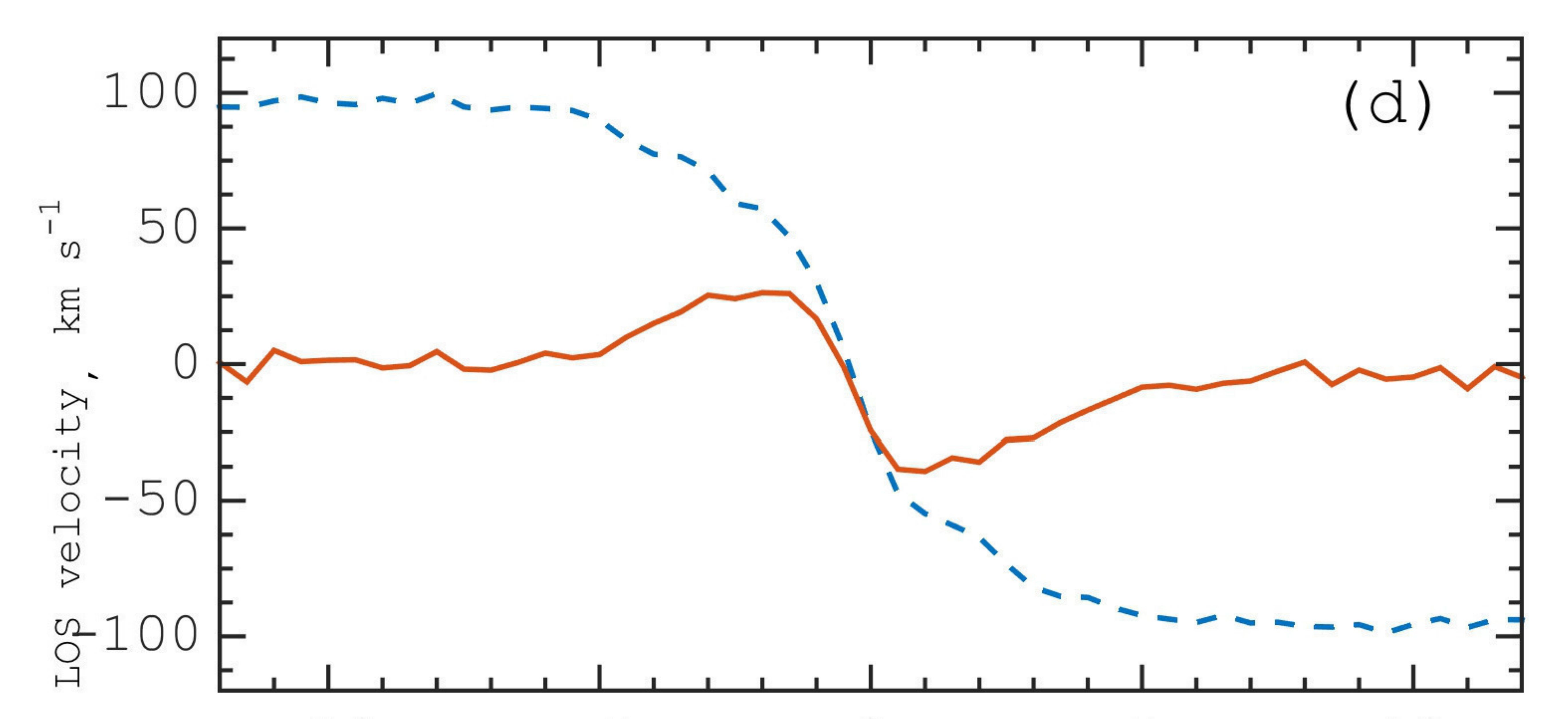}\\\includegraphics[width=0.5\hsize]{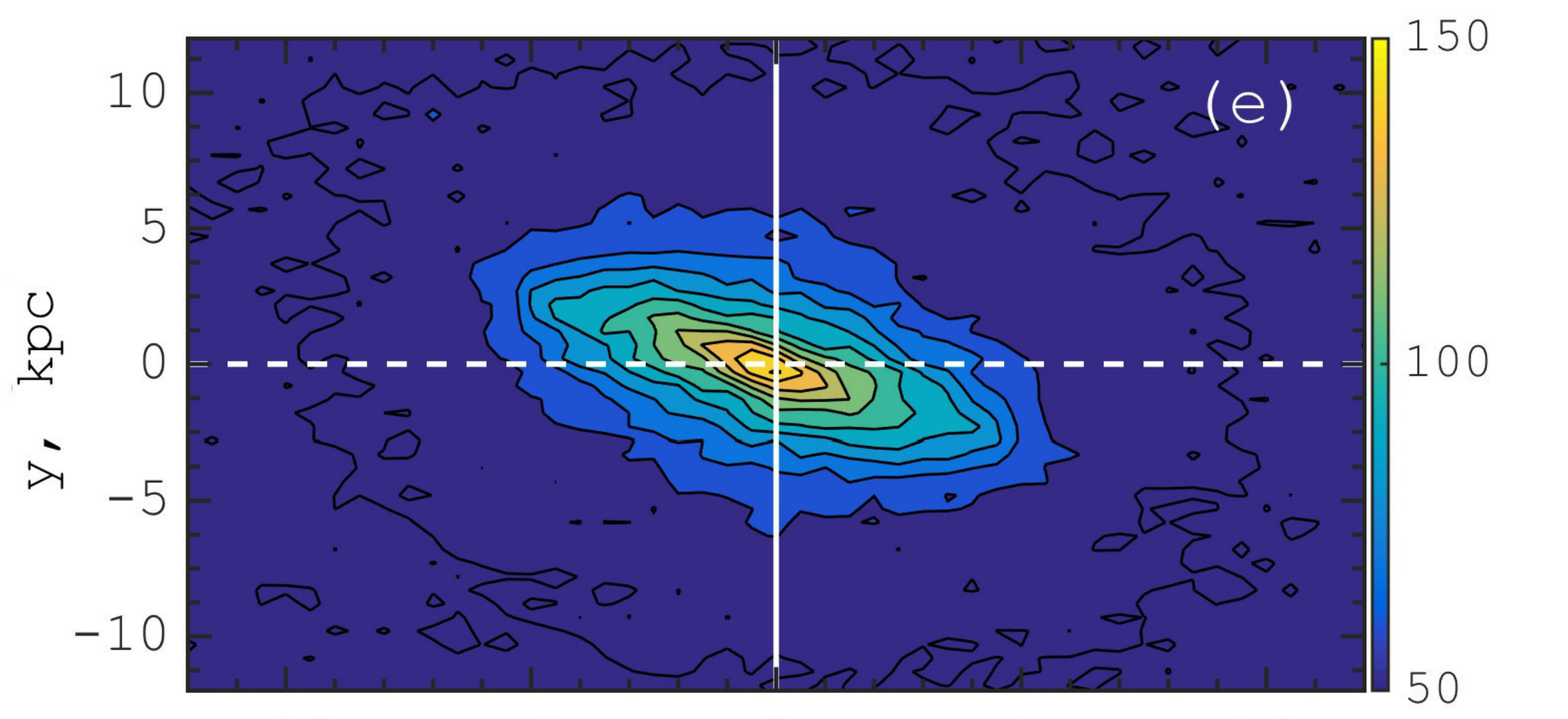}\includegraphics[width=0.5\hsize]{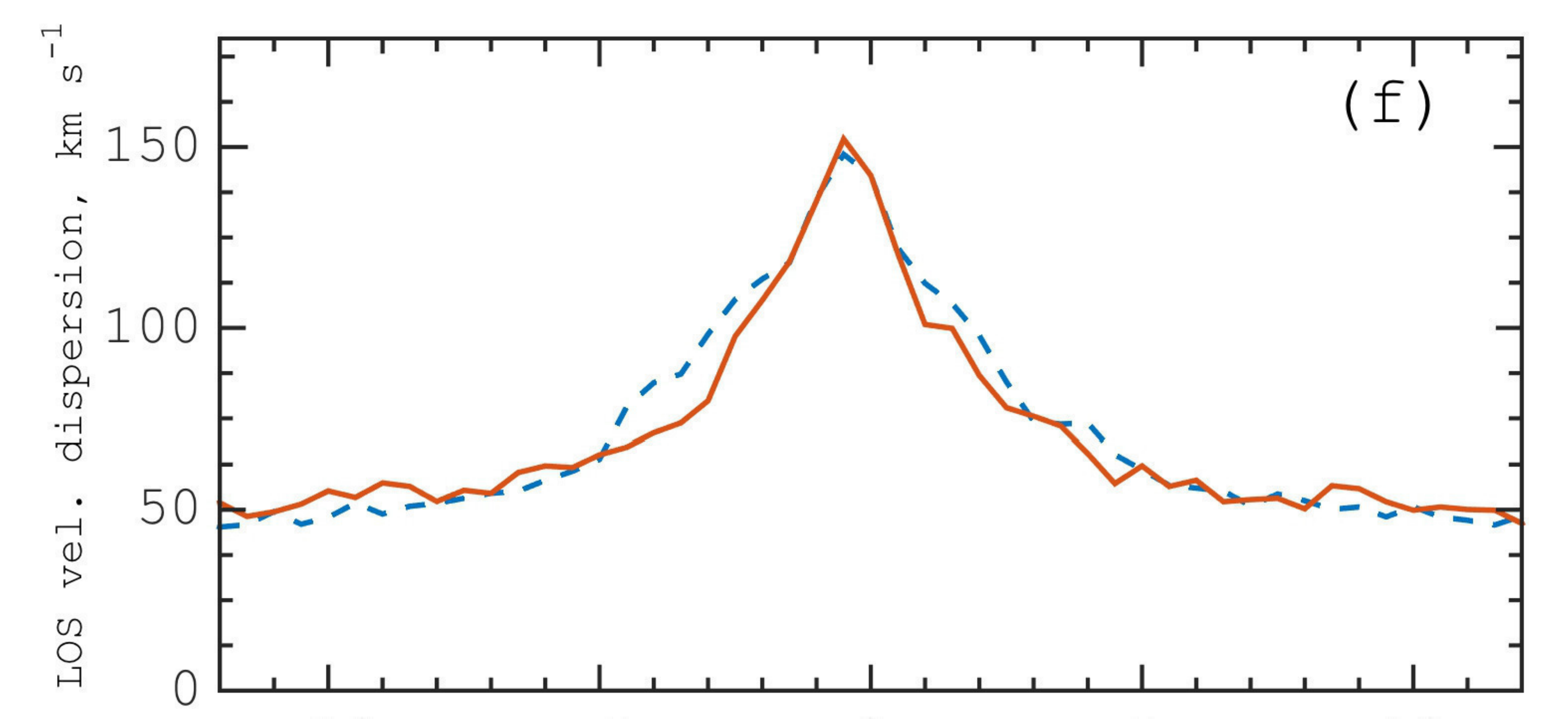}\\\includegraphics[width=0.5\hsize]{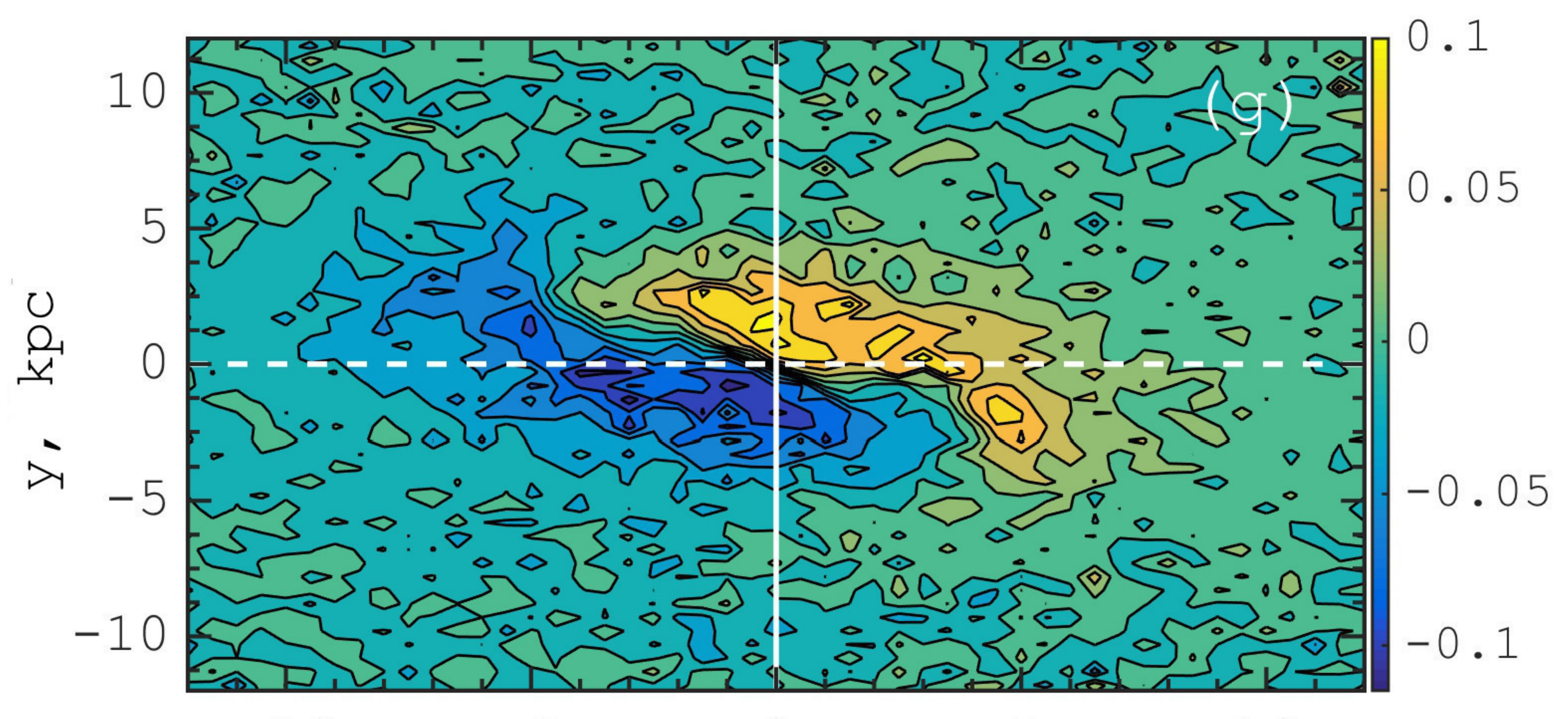}\includegraphics[width=0.5\hsize]{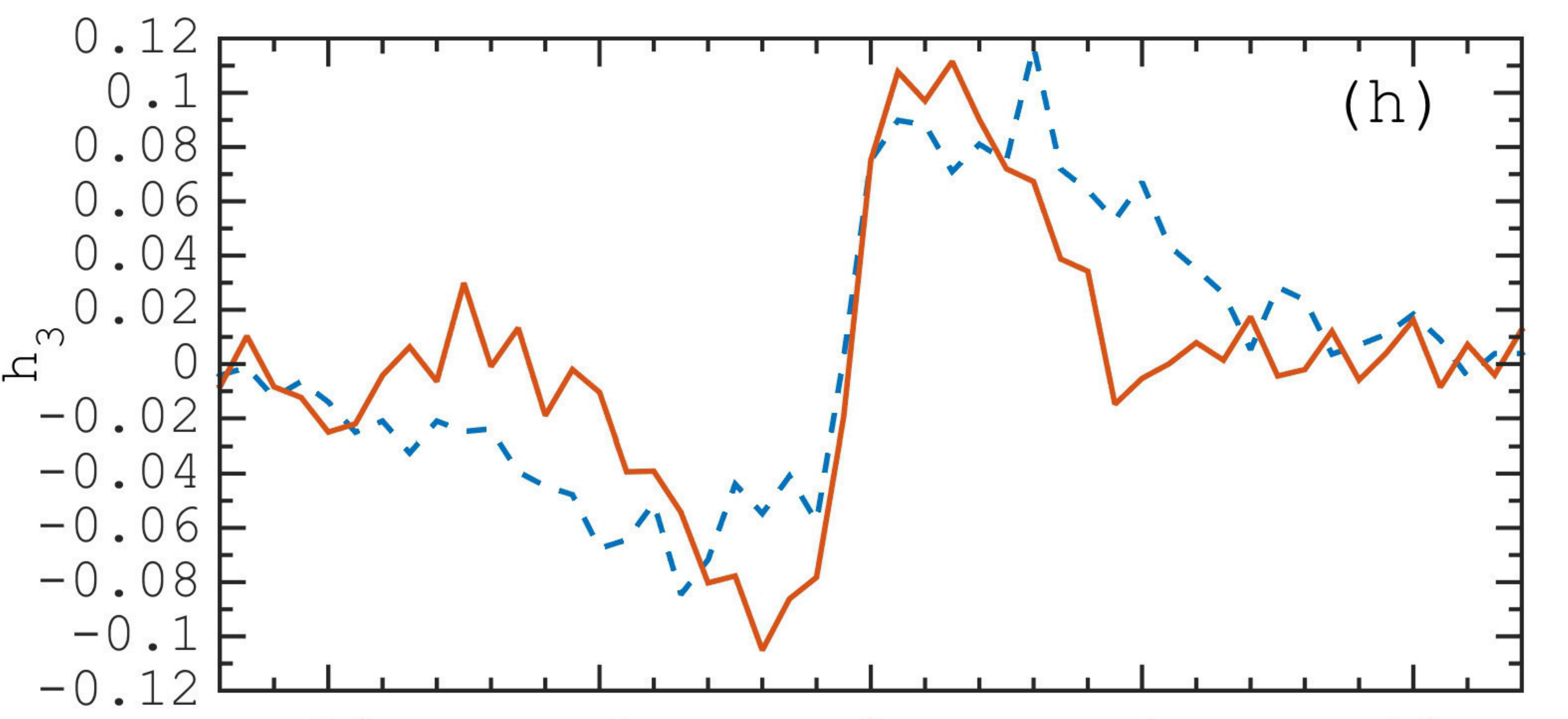}\\\includegraphics[width=0.5\hsize]{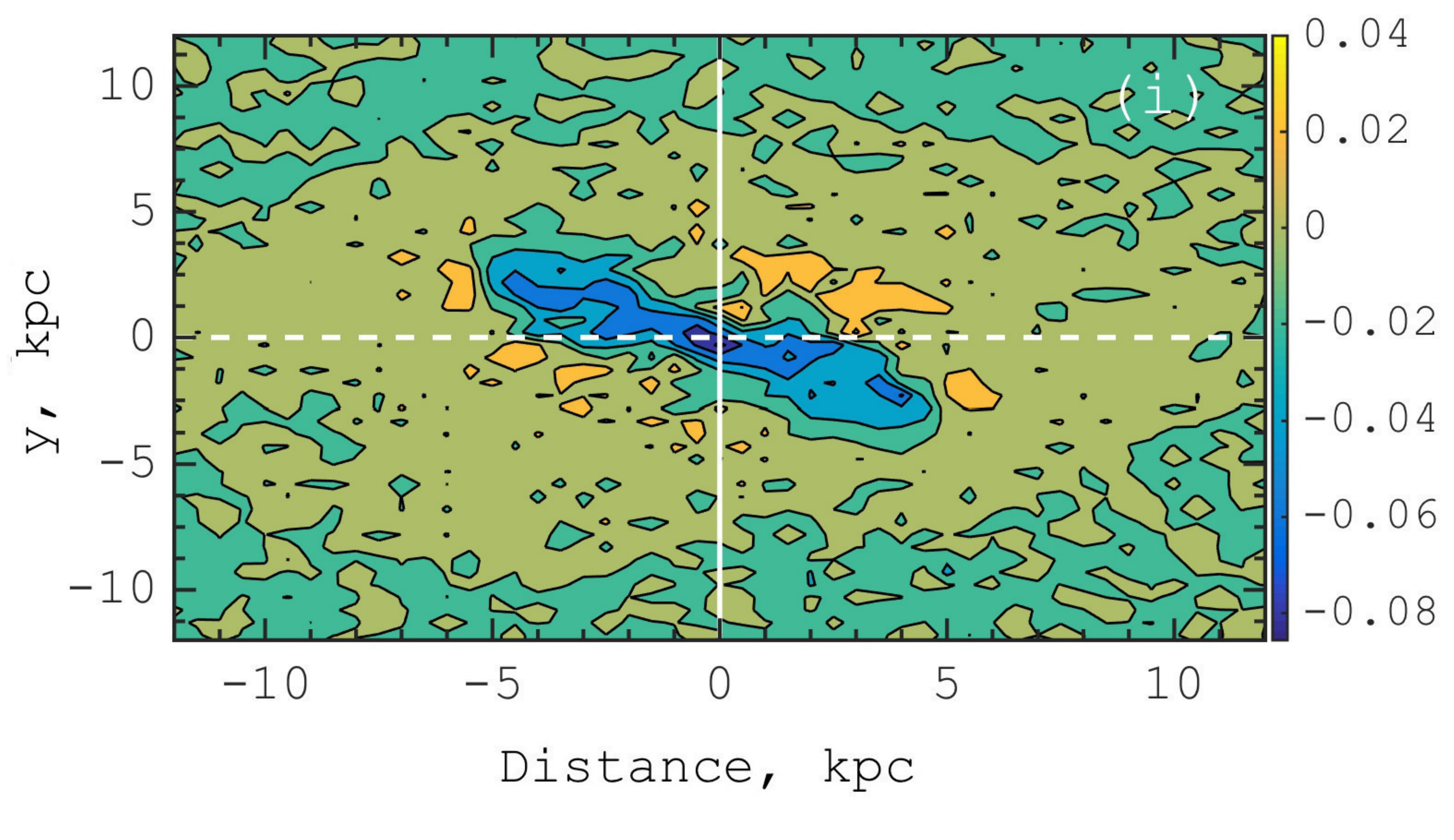}\includegraphics[width=0.5\hsize]{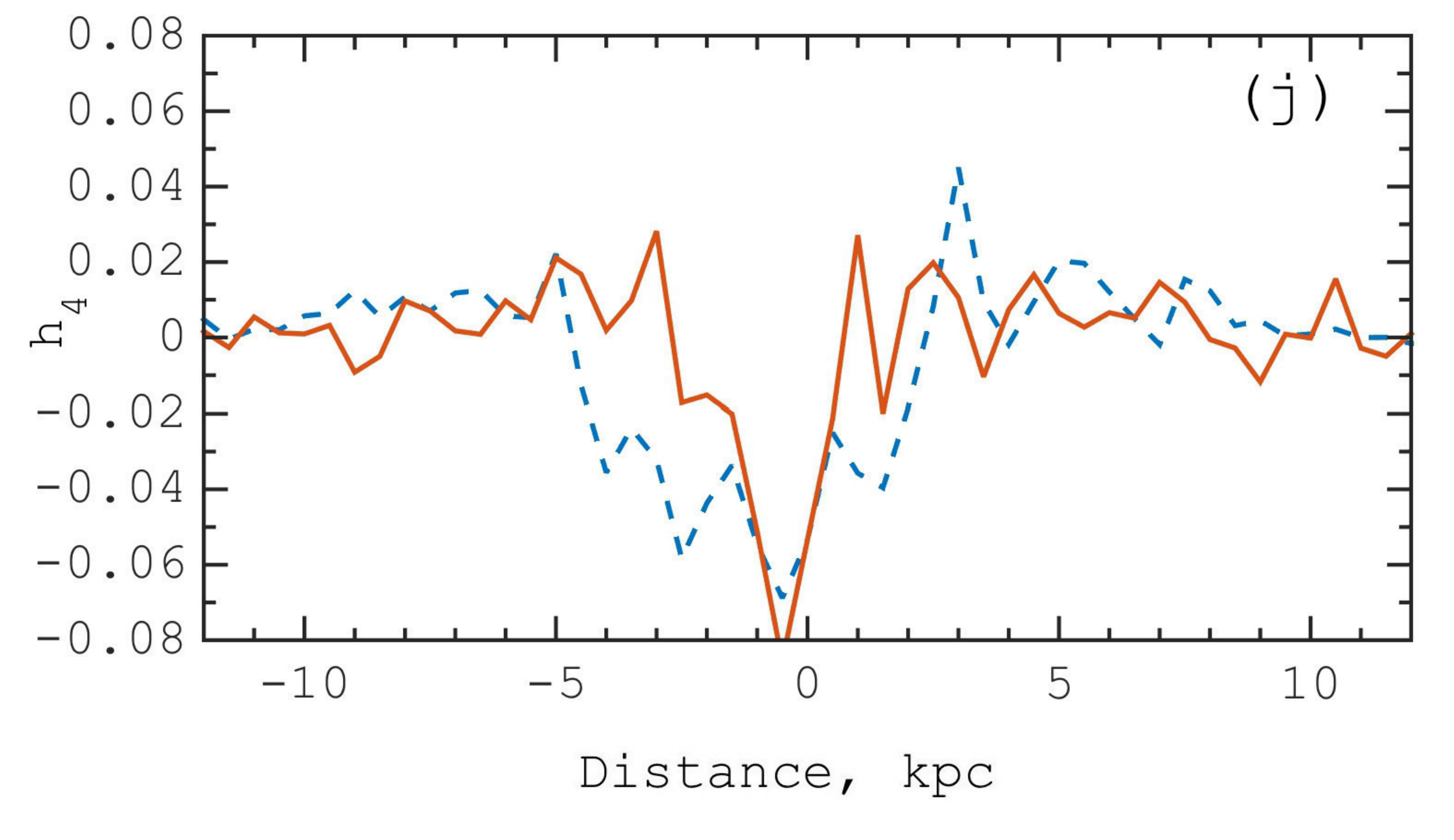}
\caption{Left column from top to bottom represents 2D maps of the stellar density, LOS velocity, LOS velocity dispersion, $h_3$ and $h_4$ of model barred galaxy respectively. The white dashed and solid lines correspond to the slit positions used to extract one-dimensional data. Bar is inclined relative to the major axis on angle $40^\circ$. Maps have been created for $500$~pc spatial resolution and $20$~\kmps resolution for the LOS velocity distribution in each spatial pixel. We adopted $30^\circ$ inclination for the simulated galaxy. In left column we show one-dimensional data along the slits shown in right. Dashed line profiles correspond to the horizontal slit~(also dashed in the left). Profiles along the vertical slit are shown by solid lines.}\label{fig::kinematics1}
\end{figure*}

\begin{figure*}
\includegraphics[width=0.5\hsize]{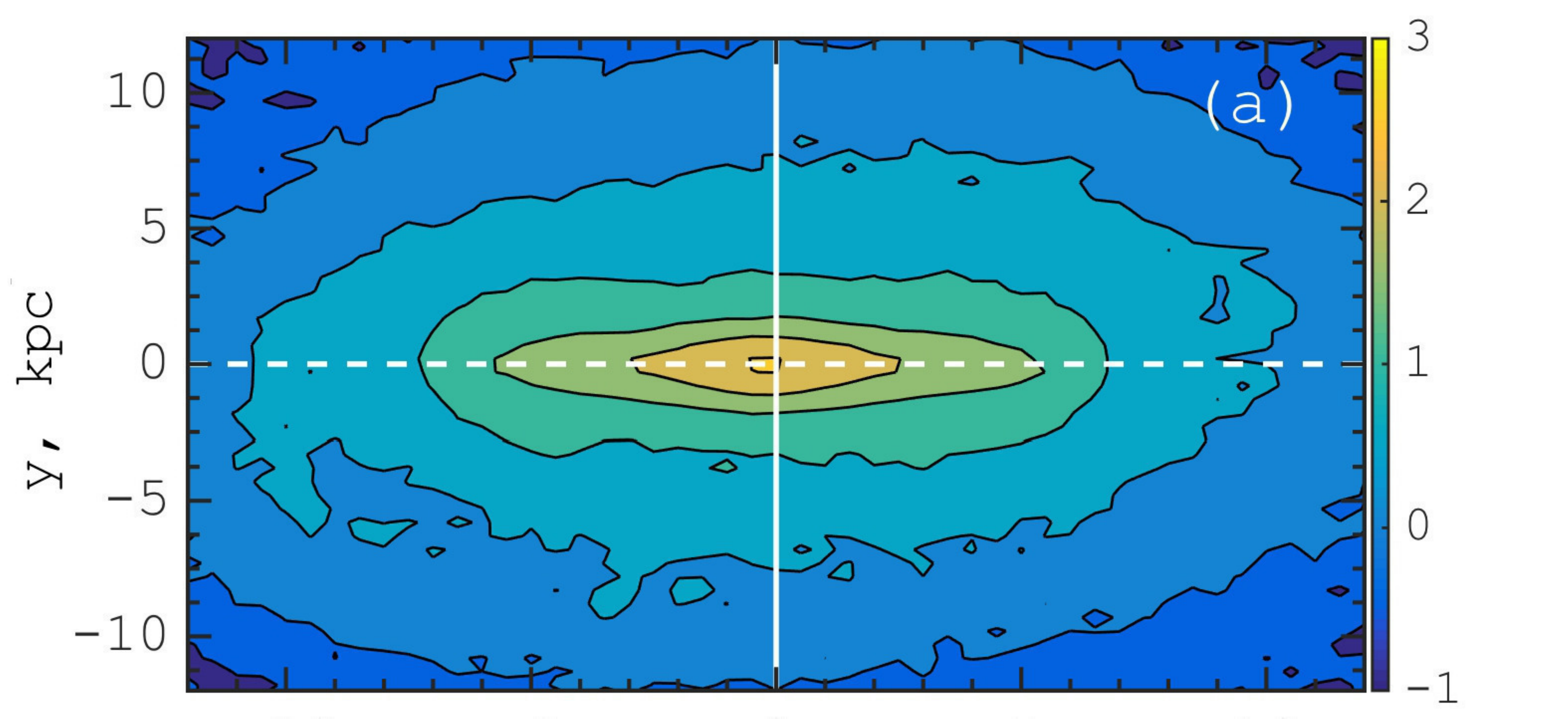}\includegraphics[width=0.5\hsize]{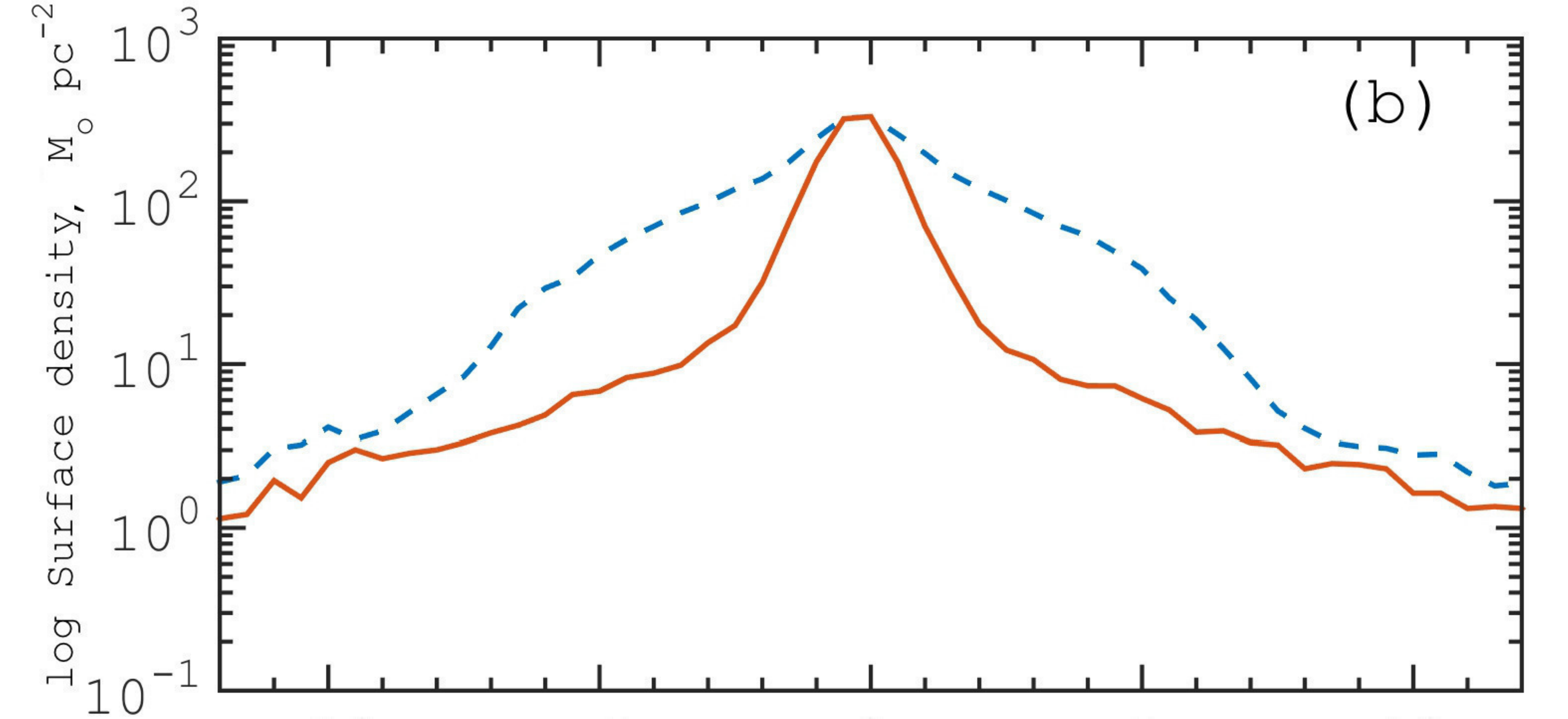}
\includegraphics[width=0.5\hsize]{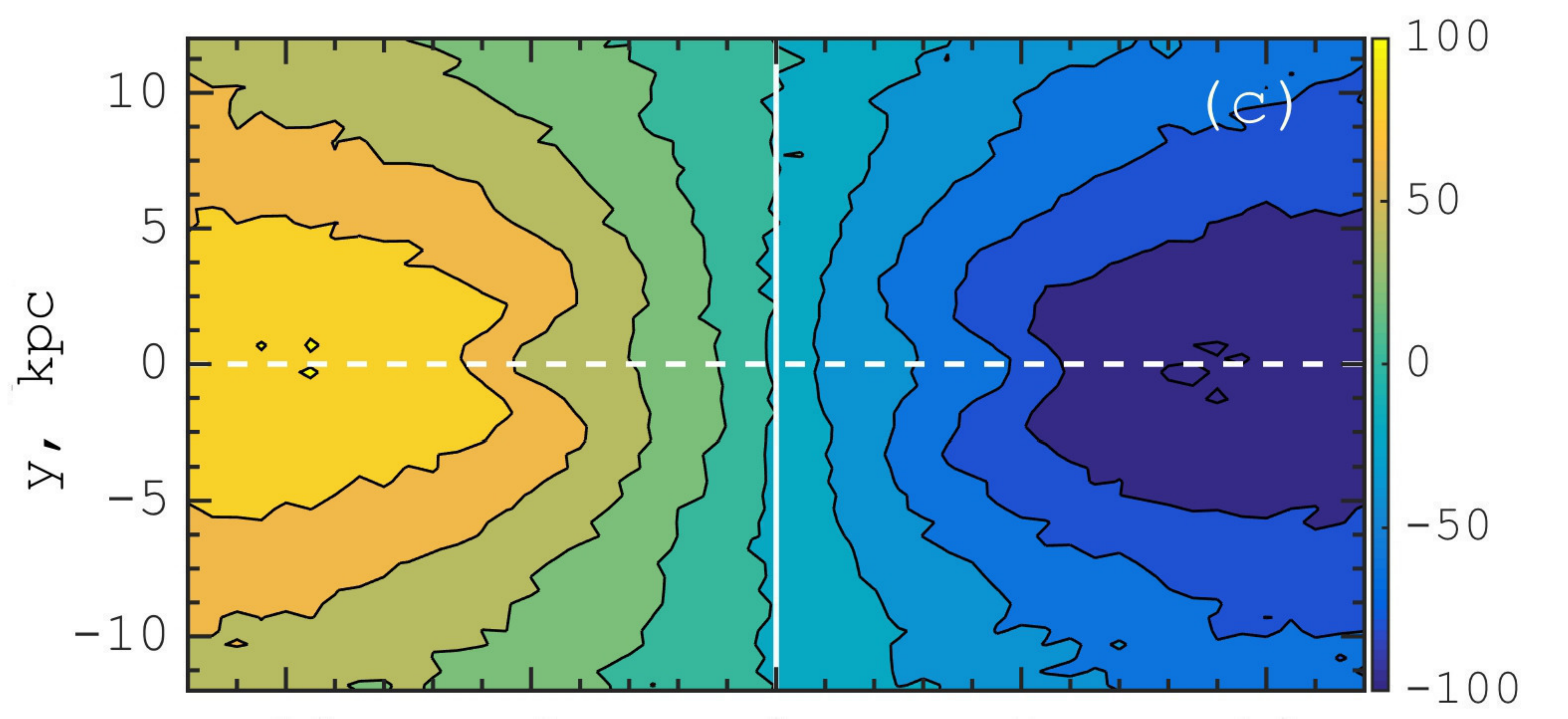}\includegraphics[width=0.5\hsize]{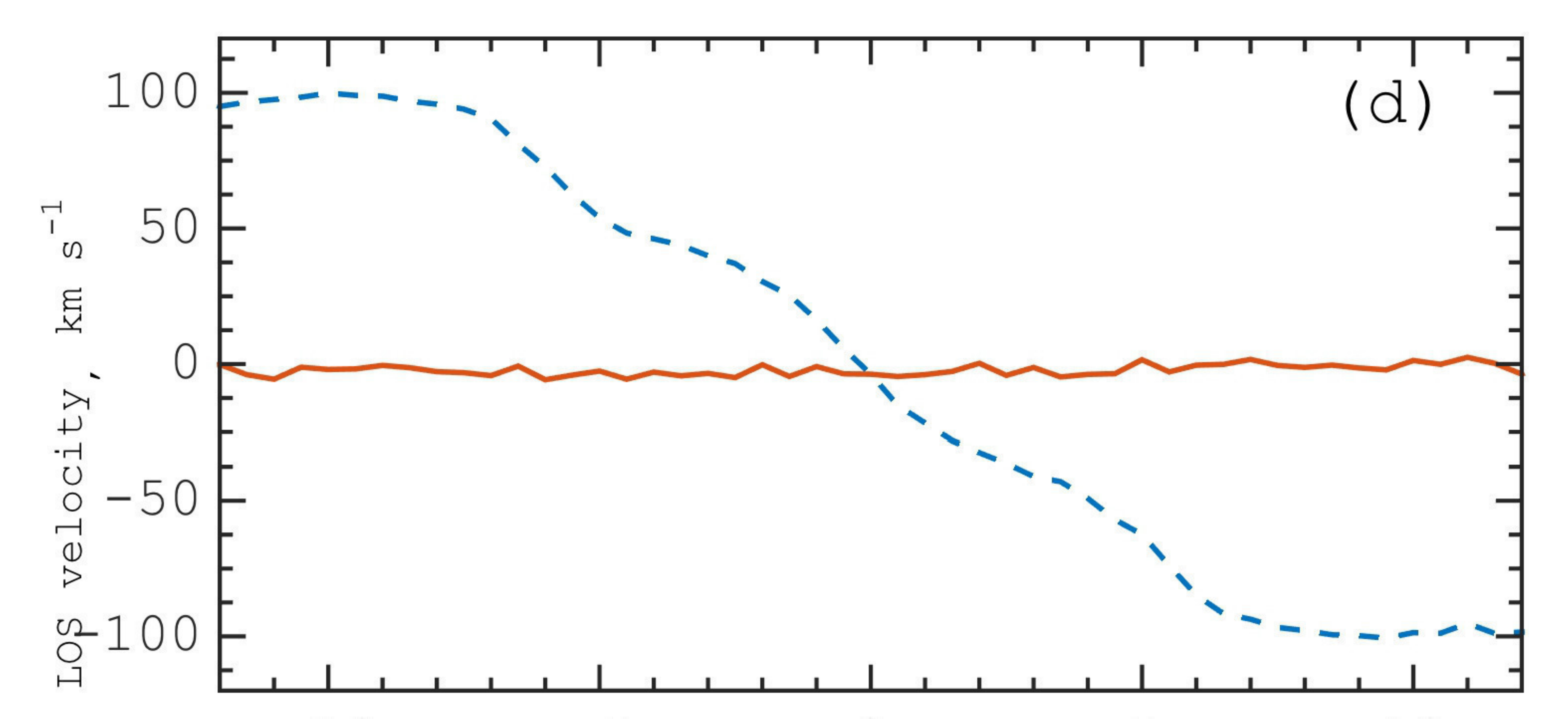}
\includegraphics[width=0.5\hsize]{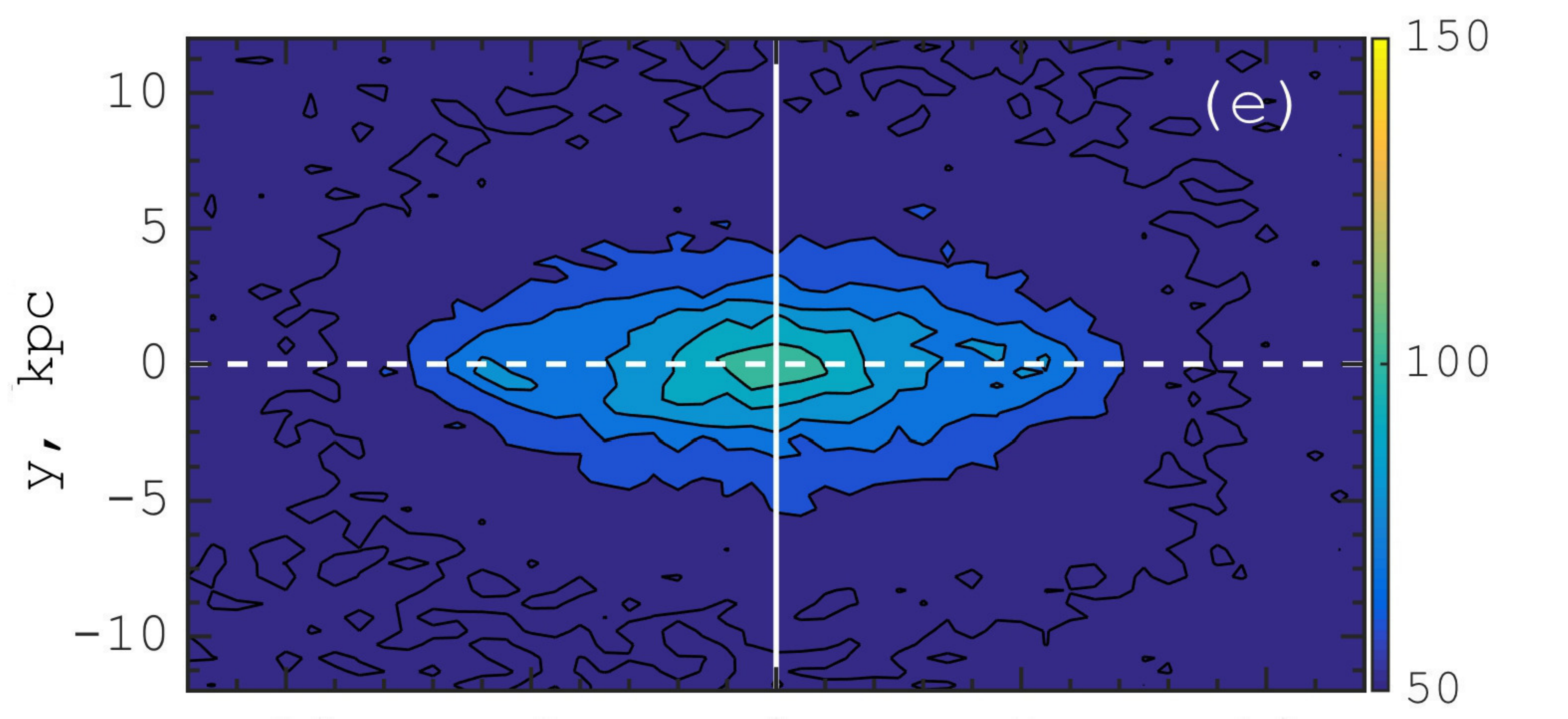}\includegraphics[width=0.5\hsize]{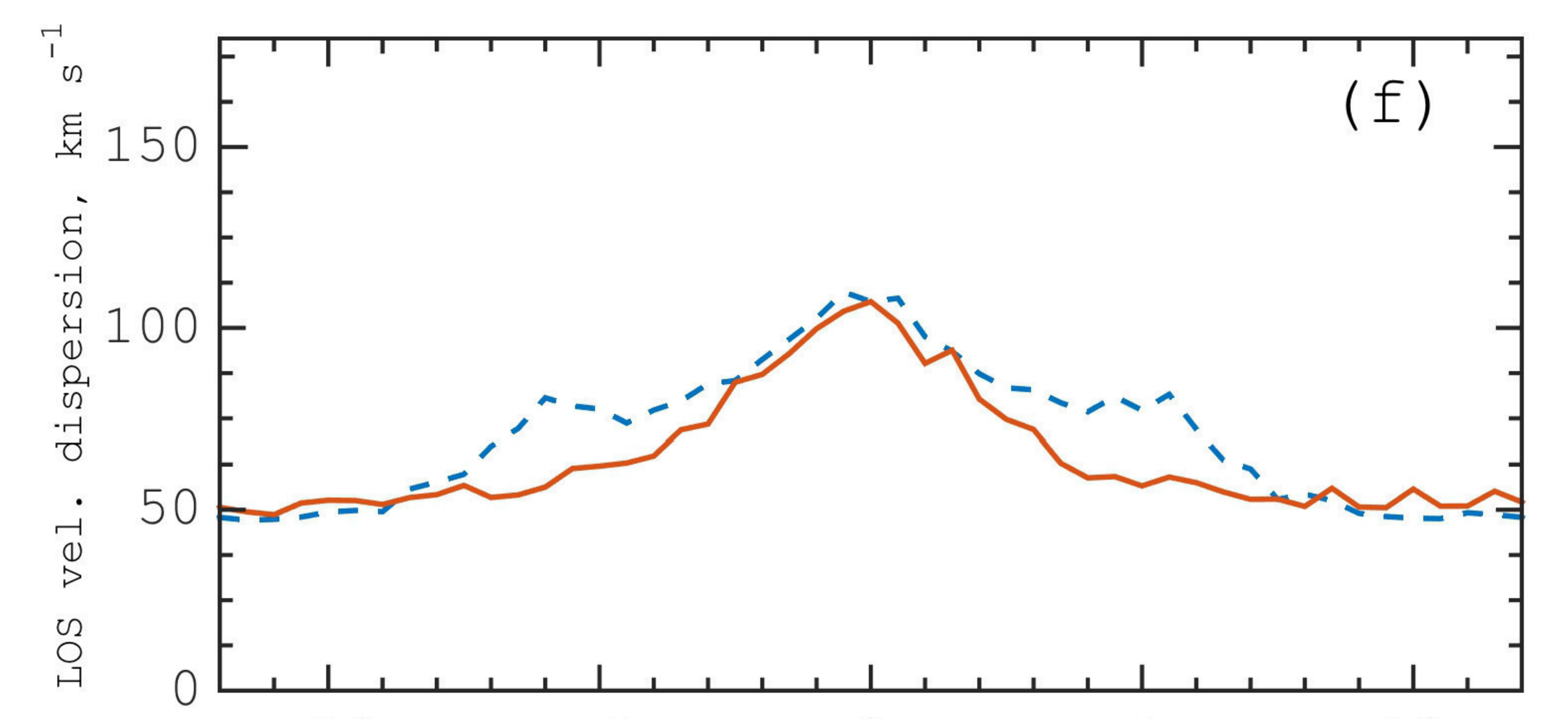}
\includegraphics[width=0.5\hsize]{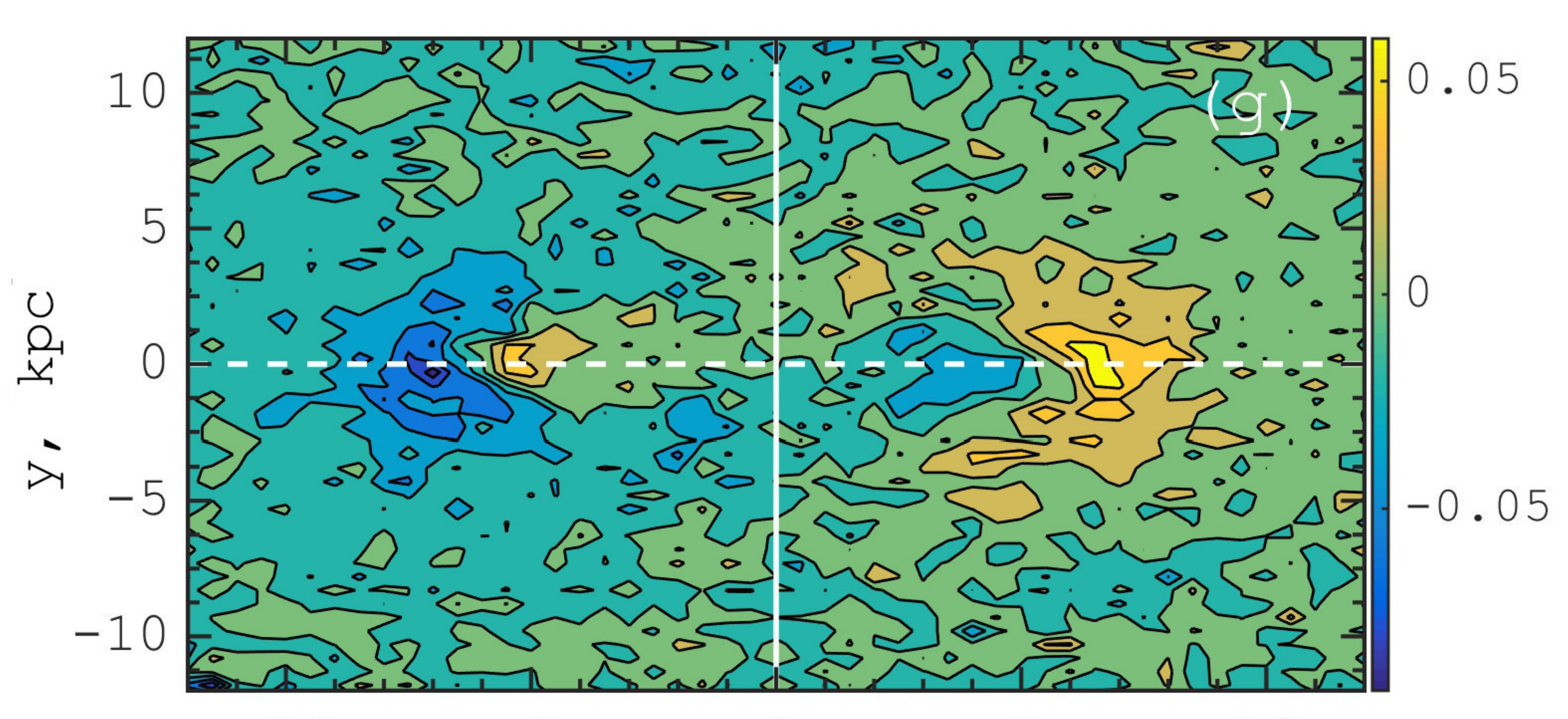}\includegraphics[width=0.5\hsize]{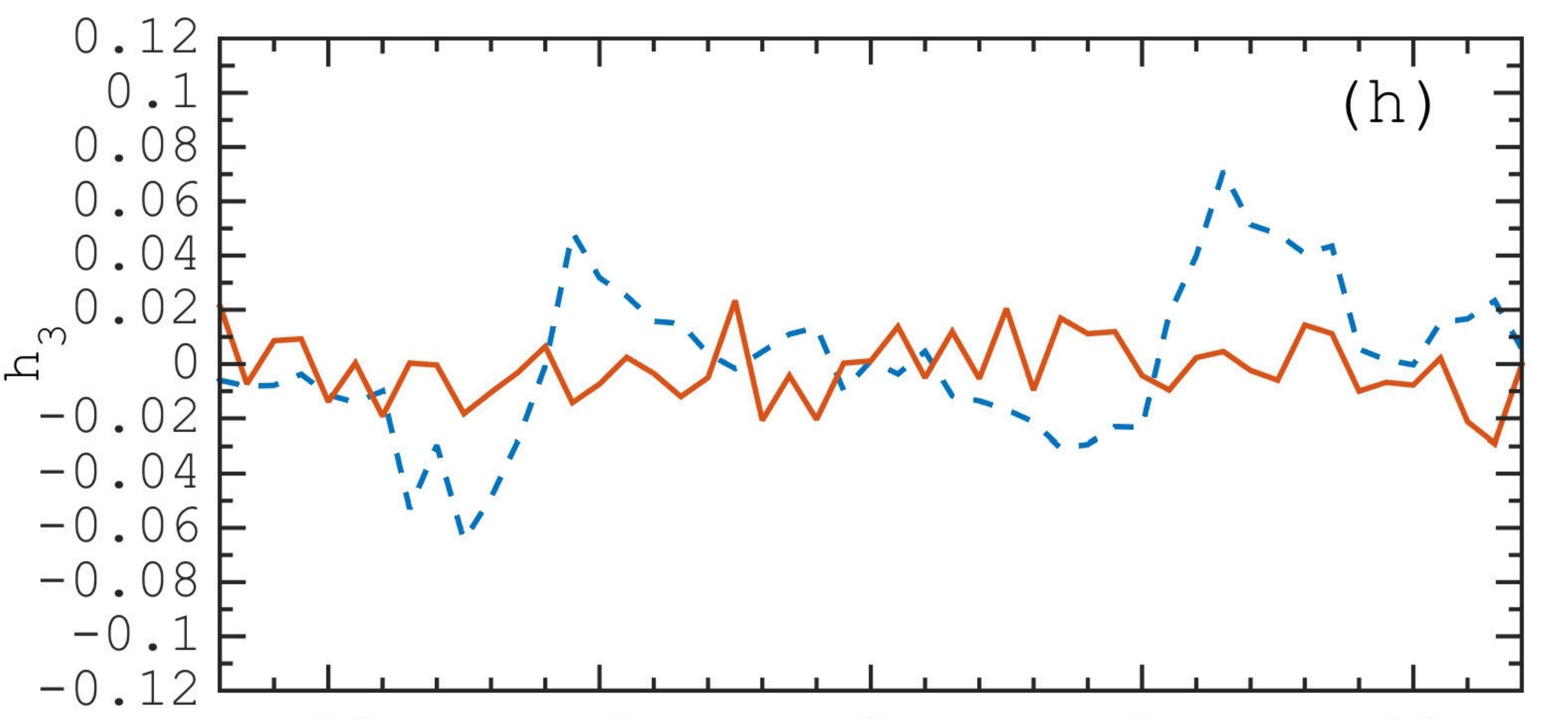}
\includegraphics[width=0.5\hsize]{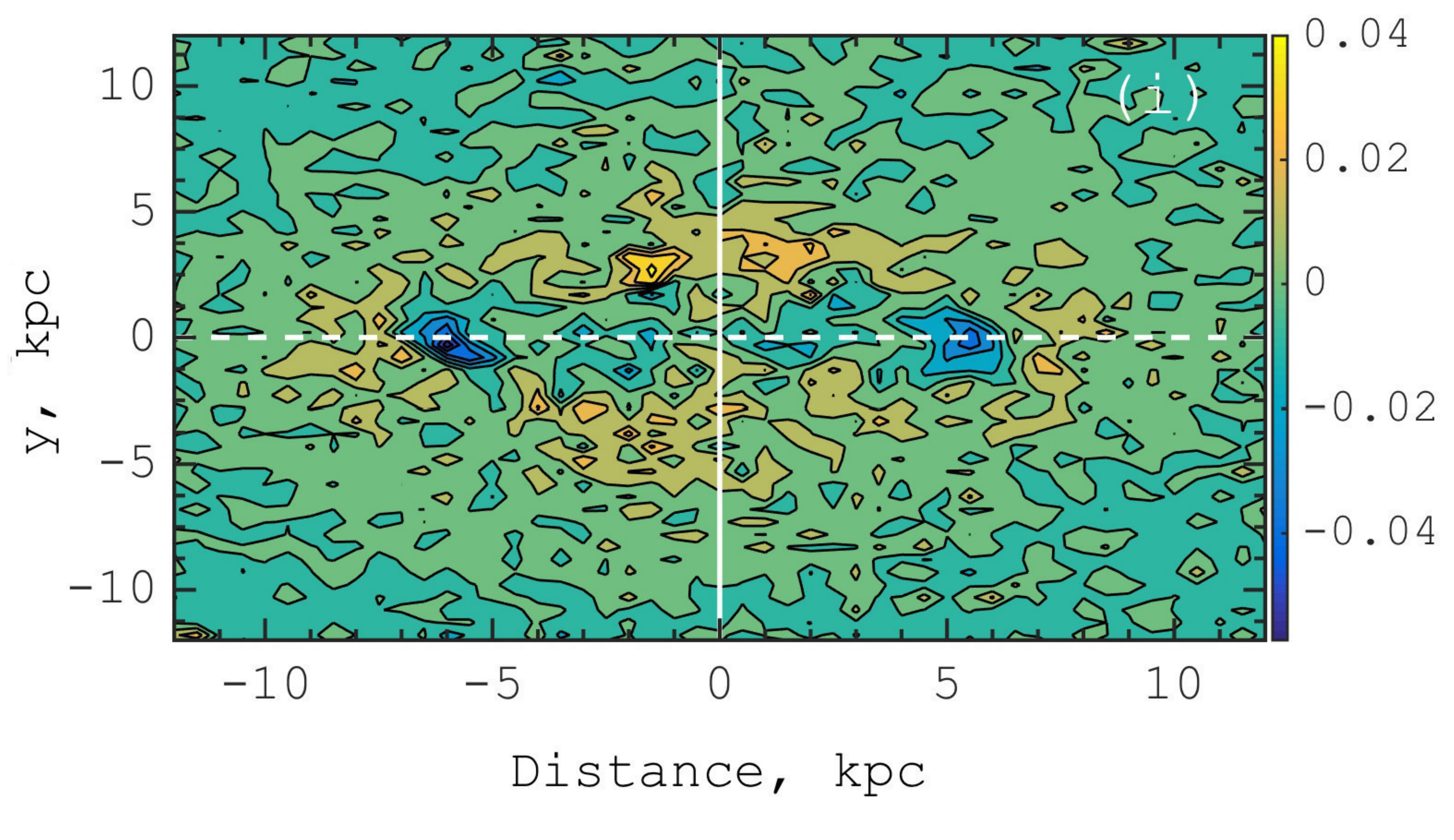}\includegraphics[width=0.5\hsize]{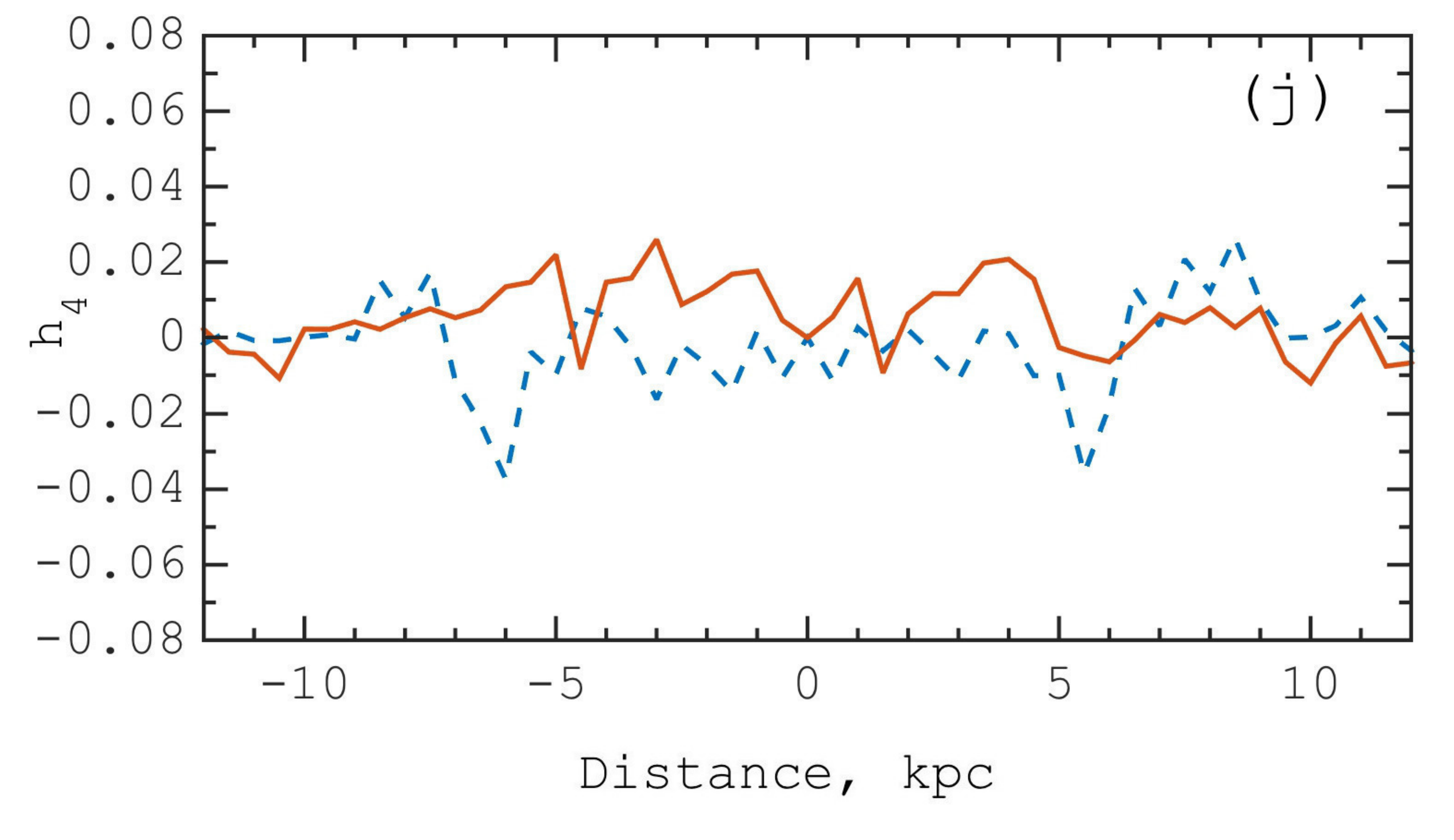}
\caption{Same as in Fig.~\ref{fig::kinematics1}, but for the bar orientated along the major axis.}\label{fig::kinematics2}
\end{figure*}

\begin{figure}
\includegraphics[width=1\hsize]{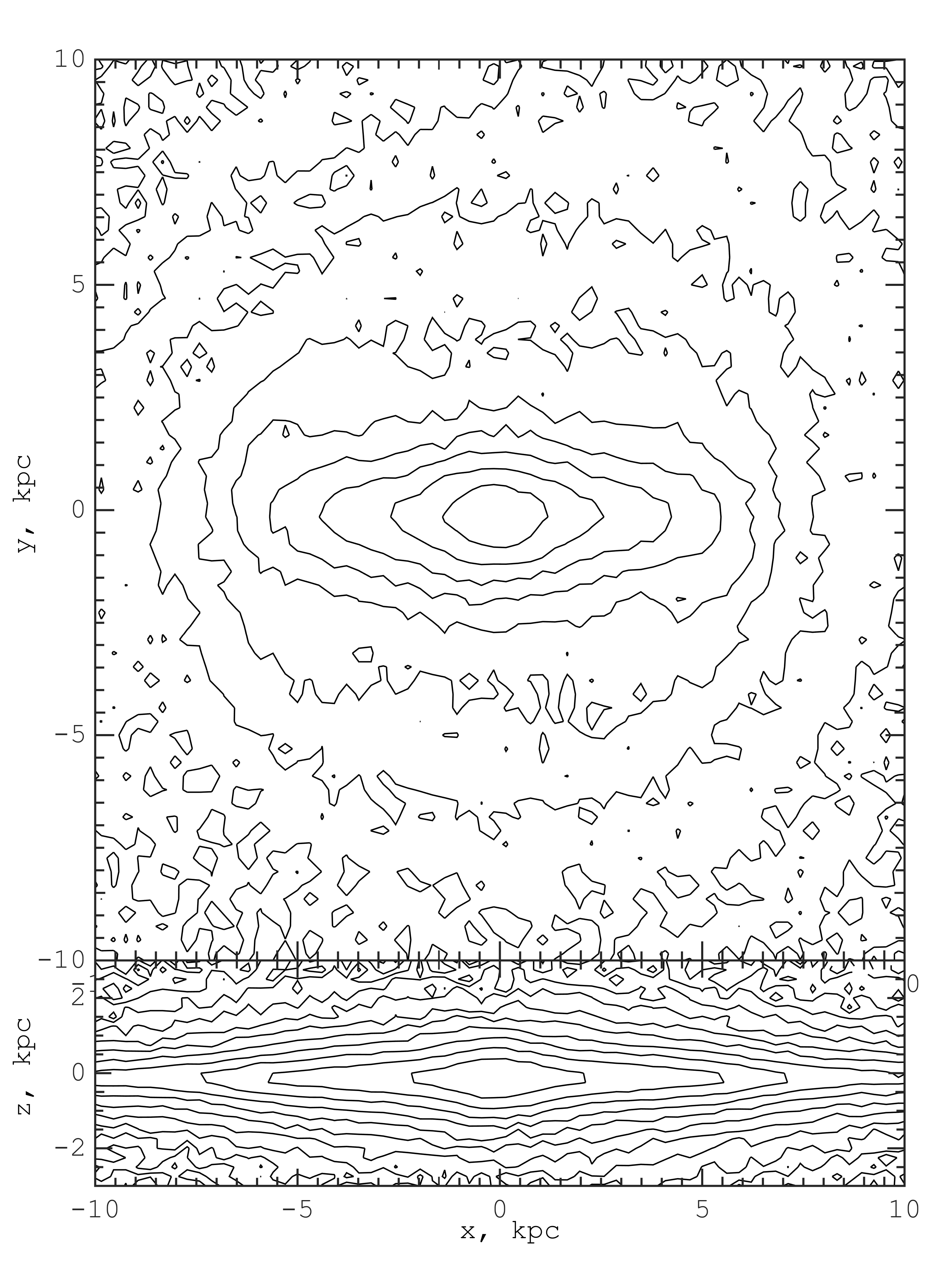}
\caption{ The  face-on (top panel) and edge-on (bottom panel) view of the density contour map of the simulated barred galaxy at 2 Gyr.}\label{projections}
\end{figure}

We compared the model profiles of velocity, velocity dispersion, $h_3$ and $h_4$ moments with the observed ones. For this purpose we created two different types of models. In the first one a bar is oriented along the major kinematical axis, in the second one it is inclined at the angle of 40\degr ~to the major axis (in the plain of a galaxy). In Figs. \ref{fig::kinematics1}, \ref{fig::kinematics2} we show the maps of the surface density, stellar velocity, velocity dispersion and $h_3$, $h_4$ moments for the models with bar inclined to the major axis and aligned with it respectively. The linear and velocity resolutions of the models are 500~pc and 20~\kmps.

For the convenience of  comparison with the observed data we performed slices of model maps of velocity, velocity dispersion and $h_3$, $h_4$  along the major (dashed lines) and minor axes (solid lines)  (see right columns of Figs. \ref{fig::kinematics1}, \ref{fig::kinematics2} for two types of bar orientation).

\subsubsection{UGC~1344}

As one can see from Figure \ref{fig::kinematics1}, the slices for the model with the bar inclined to the major axis show excellent agreement with that obtained for UGC~1344. The line-of-sight velocity profile shows similar changes of gradient. The model $h_3$ radial profile anticorrelates with that of the velocity in the centre as the observed one.  It is worth noting, however, that the observed behavior of $h_3$ profile is more complex in the innermost region than in the model. It could be due to the absence of massive boxy bulge in our model (see the face-on and edge-on view of the density countours of the model at 2 Gyr in top and bottom panels of Fig. \ref{projections} respectively). The stellar velocity dispersion profile shows the flattening at large radial distance which resembles the observed one, being however more prominent. It is worth noting that the double-humped rotation curve and a shoulder in the velocity dispersion profile were also reproduced in previous papers on $N$-body simulations of barred galaxies (see e.g. \citealt{BureauAthanassoula2005}). 

Our model $h_4$ map looks not very similar to observations~(see  Fig.  \ref{fig::kinematics1} (j) and Fig.~\ref{profiles_u1344}) -- the observed profile does not possess such a remarkable central minimum as it is seen in the model. However, the overall model behavior resembles that of UGC~1344, where the major kinematical axis is inclined with respect to the bar, as it is apparent from the kinematical profiles. We conclude that the observed kinematical features of UGC~1344 including the anticorrelation of $h_3$ and the velocity in the central region can be explained by the presence of the bar inclined with respect to the major axis without  introducing additional nuclear ring or disc, which is in a good agreement with the conclusions done in Sect~\ref{s_nonpar}. We  suspect that the inner structure represents the peanut-shaped bulge which is a result of the vertical  buckling instability of the bar during the secular evolution phase that has been suggested by a number of $N$-body simulations~\citep{1990A&A...233...82C,1991Natur.352..411R,2006ApJ...637..214M}. This conclusion follows from the stellar properties of  the inner structure not differing from the surrounding bar, its relatively high velocity dispersion and from the fact that kinematical features of UGC~1344 are in good agreement with that obtained for boxy or peanut-shaped bulges by \cite{ChungBureau2004} concerning the shape of the velocity profile and the inner anti-correlation of $h_3$ and the LOS velocity.

\subsubsection{NGC~5347}
The situation with NGC~5347 is quite different. One should notice that even though the two considered galaxies have many common features (similar kinematical behavior, the presence of a bar and a ring, the red colour of stellar population), there are some differences that could indicate the dissimilarity of formation history of these systems. First, NGC~5347 unlike UGC~1344 demonstrates the central minimum of the stellar velocity dispersion. Secondly, the age of the central component of NGC~5347 is significantly younger than that of the bar and bulge, while for UGC~1344 the central part consists of old stars. The activity of star formation throughout the entire discs of two galaxies is also different.   Based on FUV images
from the GALEX All-Sky Imaging Survey (AIS) we found out that UGC~1344 is bright in FUV only in the central region with a radius of $\sim 5$~arcsec, while the disc and the ring are almost invisible in FUV. At the same time NGC~5347 has a FUV-bright ring, indicating active star formation in the disc.

As one can see from the kinematical profiles of NGC~5347, the bar is most likely oriented along the major axis,  in contrast to UGC~1344. A comparison with the model with the same bar orientation (see Fig. \ref{fig::kinematics2}) does not show a good concordance. The model velocity profile demonstrates the change of the gradient in agreement with observations, although it is not so sharp as observed. The $h_3$ profile for the model does not demonstrate the anticorrelation with the velocity in the central part, instead the model $h_3$ profile shows slight positive trend with LOS velocity, although the values of $h_3$ are close to zero. The difference in the behavior of model $h_3$ horizontal cuts for the bar aligned along the major axis of the disc and inclined to it  can be explained as follows. The horizontal cut for the model with the bar inclined to the major axis crosses the bar region in central $\pm2$ kpc, where one can see the dramatic change of $h_3$, which anticorrelates with LOS velocity, since the bar rotates slower than the part of stars of the disc with circular orbits. The contribution of the bar to the surface brightness is very high  in this region, so this detail is very prominent on the profile. For the model with the bar elongated along the major axis of the disc the cut is oriented along the bar, so the stars belonging to the bar make the major contribution to the luminosity and thus to the line profiles.  Some  fraction of stars does not belong to the bar and thus rotates faster which seeks to provide  a positive trend of $h_3$ with the LOS velocity along the cut.  However this component is faint in comparison to the bar, which results in much less dramatic changes of the $h_3$ profile.   The model velocity dispersion profile also does not have such prominent central minimum as we found for NGC~5347. Thus, our simple barred model can not reproduce the kinematical features of NGC~5347. We compared our profiles for this galaxy with the results of high resolution simulation of a galaxy forming out of gas which cools and settles into a disc, considered by \cite{Cole2014}. They studied a model galaxy with a nuclear disc formed by gas which was driven by bar to the centre. According to their model maps there is a minimum of the stellar age and velocity dispersion in the centre and a slight peak of $h_4$  which postulates the presence of a kinematically cold stellar population. Similar drop of the stellar velocity dispersion was reproduced also in \cite{Wozniak2003}, where the young population dominates the observed line-of-sight kinematics of the circumnuclear regions. A number of numerical simulations suggest that bars induce torques which force the gas to shock, subsequently losing angular momentum and funneling to the centre~\citep{1992MNRAS.259..345A,2004ApJ...600..595R,2012ApJ...758...14K}. Such gas inflow is stabilized by  inner Lindblad resonance creating star forming nuclear disc~\citep{Sheth2005,2011MNRAS.416.2182E}. There are also observational evidences that presence of bar and bar-driven inflow leads to rejuvenation of stellar population in the nuclear region~\citep{Moorthy2006, Coelho2011, Perez2011, Sil2011, Lin2016}. This picture agrees with our findings for NGC~5347 and indicates the presence of kinematically cold nuclear disc formed through the bar-driven inflow and the following star formation.

\section{Conclusions}\label{conclusion} 

The first result of this work is the disproof of the previously suspected peculiarly low stellar mass-to-light ratios of  NGC~5347 and UGC~1344, adding evidences in favour of the universality of the IMF in discy galaxies.. In both galaxies the rotational velocities obtained from the linewidth of H{\sc i} and presented in Hyperleda appear to be strongly underestimated. We found the following new estimates of mass-to-light ratios within the optical radii: for UGC~1344 $M/L_B=3.75$ ($M/L_B<0.34$ according to H{\sc i} linewidth measurements); for NGC~5347 $M/L_R=2.3$ ($M/L_R=0.65$ according to previous  velocity measurements).  However, our long-slit spectral observations reveal some curious features in their kinematical profiles  which we tried to reproduce in our $N$-body simulations of barred galaxies. Both galaxies have double-humped rotation curves within several kpc from the centres and the central  components, probably connected with the presence of a bar. 

These  two  galaxies look quite similar at first sight: both are early-type discy galaxies possessing  a bar and inner ring. However NGC~5347 has younger stellar population and lower luminosity in comparison to UGC~1344. FUV images reveal that star formation is active throughout the disc of NGC~5347 while UGC~1344 is noticeable in FUV only in the central region.   The presence of bars evidently has played a significant role in the evolutions of central parts of both galaxies, but in a different way. In UGC~1344 the bar led to the formation of  inner component with high velocity dispersion and low rotation velocity but with the age and metallicity close to that of the bar. This central component  most probably is a peanut-shaped bulge. In NGC~5347 the bar induced  the formation of the other type of structure -- a  kinematically cold nuclear disc. Its presence is evident from the central minima on the stellar velocity dispersion and the stellar age profiles and the  anti-correlation between $h_3$ moment and velocity which are not reproduced by our $N$-body model with similar orientation of the bar, but agrees with the simulation of a galaxy with nuclear disc of \cite{Cole2014}. 

We found that stellar abundancies in the central parts of the galaxies have similar nearly solar values in spite of their different luminosities. Both galaxies demonstrate the flattening of the stellar metallicity gradient in the bar regions, although their radial profiles of abundancy and  mean stellar age are very different, evidencing a different history of star formation in these barred galaxies.

\section*{Acknowledgements}  
We thank the anonymous referee for helpful comments.  
Authors thank Daniela Bettoni for useful discussion and comments.
The Russian Science Foundation (RSCF) grant No. 14-22-00041 supported the study of stellar and ionized gas kinematics. Authors are also grateful to the RFBR grants No. 15-52-15050, No. 16-02-00649 and No. 16-32-60043 for support. The project used computational resources at the Research Computing Center funded by the M.V.~Lomonosov Moscow State University Program of Development. The Russian 6-m telescope is exploited under the financial support by the Russian Federation Ministry of Education and Science (agreement No14.619.21.0004, project ID RFMEFI61914X0004). This work has been supported by the ANR (Agence Na- tionale de la Recherche) through the MOD4Gaia project (ANR-15-CE31-0007, P.I.: P. Di Matteo). This research has made use of the Lyon Extragalactic Database (LEDA, \url{http://leda.univ-lyon1.fr}). In this study, we used the SDSS DR13 data. Funding for the SDSS and SDSS-II has been provided by the Alfred P. Sloan Foundation, the Participating Institutions, the National Science Foundation, the U.S. Department of Energy, the National Aeronautics and Space Administration, the Japanese Monbukagakusho, the Max Planck Society, and the Higher Education Funding Council for England. The SDSS
Web site is \url{http://www.sdss.org/}.

\bibliographystyle{mnras}
\bibliography{Saburova}

\label{lastpage}

\end{document}